\documentclass[12pt]{article}
\pdfoutput=1
\usepackage{color}
\usepackage{amssymb,amsmath}
\usepackage{epsf}
\usepackage{epsfig}
\usepackage{afterpage}
\usepackage{longtable}
\usepackage{cite}
\usepackage{latexsym,mathrsfs}
\usepackage{graphics}
\usepackage{url}
\usepackage{paralist}
\usepackage{caption}
\newcommand{\e}{\mathrm{e}}

\newcommand{\ord}{\mathcal{O}}

\newcommand{\tev}{\, {\rm TeV}}
\newcommand{\gev}{\, {\rm GeV}}
\newcommand{\mev}{\, {\rm MeV}}

\newcommand{\vcb}{|V_{cb}|}

\newcommand{\ts}{\tilde s}
\newcommand{\tc}{\tilde c}

\newcommand{\eq}[1]{(\ref{#1})}

\def\epe{\varepsilon'/\varepsilon}
\newcommand{\beq}{\begin{equation}}
\newcommand{\eeq}{\end{equation}}
\newcommand{\be}{\begin{equation}}
\newcommand{\ee}{\end{equation}}
\newcommand{\bi}{\begin{itemize}}
\newcommand{\ei}{\end{itemize}}
\newcommand{\ba}{\begin{array}}
\newcommand{\ea}{\end{array}}
\newcommand{\beqa}{\begin{eqnarray}}
\newcommand{\eeqa}{\end{eqnarray}}
\newcommand{\bea}{\begin{eqnarray}}
\newcommand{\eea}{\end{eqnarray}}
\newcommand{\beqn}{\begin{eqnarray}}
\newcommand{\eeqn}{\end{eqnarray}}

\newcommand{\eps}{\epsilon}
\newcommand{\nn}{\nonumber}

\definecolor{red}{cmyk}{0,1,1,0.4}

\def\kpn{K^+\rightarrow\pi^+\nu\bar\nu}
\def\klpn{K_{L}\rightarrow\pi^0\nu\bar\nu}

\usepackage{fancyhdr}
\advance \headheight by 3.0truept       

\begin{document}

\begin{flushright}
    {AJB-19-2}\\
    {BARI-TH/19-722}
\end{flushright}

\medskip

\begin{center}
{\LARGE\bf
\boldmath{Quark-Lepton Connections in $Z^\prime$ Mediated FCNC Processes: Gauge Anomaly Cancellations at Work}}
\\[0.8 cm]
{\bf Jason~Aebischer$^{a}$,  Andrzej~J.~Buras$^{b}$,\\
Maria Cerd{\`a}-Sevilla$^{c}$ and  Fulvia~De~Fazio$^{d}$
 \\[0.5 cm]}
{\small
$^a$ Department of Physics, University of California at San Diego, La Jolla, CA 92093, USA \\
$^b$TUM Institute for Advanced Study, Lichtenbergstr. 2a, D-85747 Garching, Germany\\
$^c$ Excellence Cluster Universe, Boltzmannstr.~2, 85748~Garching, Germany \\
$^d$Istituto Nazionale di Fisica Nucleare, Sezione di Bari, Via Orabona 4,
I-70126 Bari, Italy}
\end{center}

\vskip0.41cm

\abstract{%
\noindent
We consider scenarios with a heavy $Z^\prime$ gauge boson with flavour
non-universal quark and lepton couplings with the goal to illustrate
how the cancellation of gauge anomalies generated by the presence of an additional
 $\text{U(1)}^\prime$  gauge symmetry would imply correlations between FCNC processes within
the quark sector, within the lepton sector and most interestingly between quark flavour
and lepton flavour violating processes. To this end we present  simple
scenarios with only left-handed flavour-violating $Z^\prime$ couplings and those
in which also right-handed flavour-violating couplings are present. The considered scenarios are characterized by a small number of free parameters but in contrast to gauge anomaly cancellation in the Standard Model, in which it takes
place separately within each generation, in our scenarios anomaly cancellation
involves simultaneously quarks and leptons of all three generations.
Our models involve, beyond the ordinary quarks and leptons, three heavy right-handed neutrinos. The
models with only left-handed FCNCs of $Z^\prime$ involve beyond
$g_{Z^\prime}$ and $M_{Z^\prime}$  two real parameters characterizing the charges of all fermions under the $\text{U(1)}^\prime$ gauge symmetry and the CKM and PMNS ones in the quark and lepton sectors, respectively. The models with the right-handed FCNCs of $Z^\prime$ involve few additional parameters. Imposing constraints
from well measured $\Delta F=2$ observables we identify a number of interesting
correlations that involve e.g. $\epe$, $B_{s,d}\to\mu^+\mu^-$, $B\to K(K^*) \ell^+\ell^-$, $\kpn$, $\klpn$ and purely lepton flavour violating
decays like $\mu\to e\gamma$, $\mu\to 3 e$, $\tau\to 3\mu$ and $\mu-e$ conversion} among others. Also $(g-2)_{\mu,e}$ are considered.  The impact of the experimental  $\mu\to e\gamma$,  $\mu\to 3 e$ and in particular $\mu-e$ conversion bounds on rare $K$ and $B$ decays is emphasized.

\thispagestyle{empty}
\newpage
\setcounter{page}{1}

\tableofcontents
\newpage

\section{Introduction}

Among the simplest new physics (NP) scenarios there are the ones  with a new heavy neutral
gauge boson $Z^\prime$ and the associated gauge symmetry $\text{U(1)}^\prime$. There
is a long history of $Z^\prime$ models with selected papers and reviews given
in \cite{Langacker:2008yv,Erler:2009jh,Buras:2012jb}. In particular, already
long time ago $Z^\prime$ models with flavour-non-universal couplings to SM
fermions have been considered \cite{Langacker:2000ju,Barger:2003hg,Barger:2004qc,Barger:2009qs,Barger:2009eq}. In the context of recently very popular
$B$-physics anomalies selected papers are given in \cite{Celis:2015ara,Altmannshofer:2014cfa}.

One of the important issues in such models is the
cancellation of the gauge anomalies that naturally are generated in the
presence of an additional $\text{U(1)}^\prime$ gauge group. This case has been studied in
\cite{Carena:2004xs,Ellis:2017tkh,DAmico:2017mtc,Alonso:2017bff,Allanach:2015gkd,Kahlhoefer:2015bea,Ekstedt:2016wyi,Ismail:2016tod,Celis:2015ara} in the context of
collider processes and $B$-physics anomalies but in most
of these papers, the exception being \cite{Celis:2015ara,Ellis:2017tkh}, flavour universality either in the lepton sector or quark sector or in both has been assumed.
But even in \cite{Ellis:2017tkh} $Z^\prime$  couplings to the first generation of quarks have been set to zero in order to avoid stringent constraints
from the Kaon sector and in most papers new complex phases have been set to zero as they were not required to explain $B$-physics anomalies.

On the other hand, Kaon physics has still a lot to offer. In particular
the ratio $\epe$ \cite{Aebischer:2019mtr}, that describes the amount of direct CP violation in $K_L\to\pi\pi$ decays relative to the indirect one, still allows for significant NP
contributions and this also applies to rare Kaon decays $\kpn$ and $\klpn$.

But in addition to FCNC processes in the quark sector, an increasing role in
the coming years will be played by purely leptonic processes like
$\mu\to e\gamma$, $\mu\to 3 e$, $\tau\to 3\mu$ and $\mu-e$ conversion among others. Also $(g-2)_{\mu,e}$ is of great interest in view of the hints for NP and
the new measurement of  $(g-2)_{\mu}$ expected soon from Fermilab.

Recently, some correlations between $B$-physics anomalies and in particular
$\tau$ decays through renormalization group effects have been pointed out
in various papers but mostly in the context of leptoquark models or
in the context of SMEFT analyses of these anomalies. Selected papers can be
found in \cite{Feruglio:2017rjo,Buttazzo:2017ixm,Aebischer:2018iyb,Aebischer:2019mlg}.

In the present paper we want to look simultaneously
at most important FCNC processes in the quark and lepton sectors in the
context of scenarios with a heavy $Z^\prime$ gauge boson with flavour
non-universal quark and lepton couplings with the goal to illustrate
how the correlations between various observables are implied not through
 SMEFT RG effects or flavour symmetries but through
 the cancellation of gauge anomalies generated through  the presence of an additional $\text{U(1)}^\prime$  gauge symmetry.

 In general $Z^\prime$ models, like those considered in  \cite{Langacker:2008yv,Erler:2009jh,Buras:2012jb}, it is assumed that gauge anomalies in question are
 canceled by some very heavy fermions whose contributions to rare decays are strongly suppressed. Without some flavour symmetries the flavour violating $Z^\prime$ couplings in different meson systems and in charged lepton decays are then independent of each other and this also applies to flavour conserving couplings. However, if the cancellation of gauge anomalies should take place without new
 particles beyond the ordinary quarks and leptons, these couplings cannot
 be arbitrary. This, as we will demonstrate in our paper, implies correlations between
 FCNC processes within
the quark sector, within the lepton sector and most interestingly between quark flavour
and lepton flavour violating processes. To this end we present  simple
scenarios with only left-handed (LH) flavour-violating $Z^\prime$ couplings and  those
in which also right-handed (RH) flavour-violating couplings are present. The considered scenarios are characterized by a small number of free parameters but in contrast to gauge anomaly cancellation in the Standard Model (SM), in which it takes
place separately within each generation, in our scenarios anomaly cancellation
involves simultaneously quarks and leptons of all three generations.

The considered model involves, besides the SM quarks and leptons, three heavy RH neutrinos. Furthermore a heavy scalar, assumed to be a singlet under the SM gauge symmetry, is needed to provide a mass to the $Z^\prime$. Scenarios with only LH FCNCs generated by the $Z^\prime$ involve besides
$g_{Z^\prime}$ and $M_{Z^\prime}$  two real parameters characterizing the  $\text{U(1)}^\prime$ charges of all fermions. Further parameters stem from the CKM- and PMNS-like matrices in the quark and lepton sector, respectively. Scenarios with the RH FCNC $Z^\prime$-couplings involve few additional parameters. Imposing constraints
from precisely measured $\Delta F=2$ observables we identify a number of interesting
correlations that involve the $B$-decays $B_{s,d}\to\mu^+\mu^-$, $B\to K(K^*) \ell^+\ell^-$, the Kaon observables $\epe$, $\kpn$, $\klpn$ and purely lepton flavour violating
decays like $\mu\to e\gamma$, $\mu\to 3 e$, $\tau\to 3\mu$ and $\mu-e$ conversion among others. Furthermore $(g-2)_{\mu,e}$ are considered.

The outline of our paper is as follows. In Section~\ref{sec:2} we define the
$Z^\prime$ couplings to SM particles that are characterized by their charges
under the $\text{U(1)}^\prime$ gauge symmetry and we list all equations
that these charges have to satisfy to ensure the absence of gauge anomalies.
In Section~\ref{sec:3} we present a simple model for gauge anomaly
cancellation that involves only a few new parameters. In Section~\ref{sec:4}
we list the observables considered by us together with the references to our
previous papers where the relevant formulae in the same notation can be found.
For charged lepton decays, $\mu-e$ conversion and $(g-2)_{\mu,e}$ we give
explicit formulae because they cannot be found there.

In Section~\ref{sec:5} we present different scenarios for the couplings
distinguishing between models with and without RH flavour violating couplings. The numerical analysis of these scenarios is presented in Section~\ref{sec:5a}. Finally, we summarize in Section~\ref{sec:conclusions}.

\boldmath
\section{Basic Gauge Anomaly Equations}\label{sec:2}
\unboldmath

\boldmath
\subsection{Basic Lagrangian for $Z^\prime$}
\unboldmath
We start with the basic Lagrangian describing the interactions of  the $Z^\prime$ gauge boson with SM fermions in the flavour basis.
Generally, $Z-Z^\prime$ mixing is present, but we will neglect it in
the present paper as it does not affect the points we want to make. Thus,
$Z^\prime$ in our paper is from the beginning in the mass-eigenstate basis.

Using the notation of \cite{Buras:2012jb}, we have then
\be\label{ZprimeL}
\mathcal{L}(Z^\prime)=  \sum_{i,j,\psi_L} \Delta^{ij}_L(Z^{\prime})\, \bar\psi_L^i\gamma^\mu P_L\psi_L^jZ^{\prime}_\mu +\sum_{i,j,\psi_R} \Delta^{ij}_R(Z^{\prime})\, \bar\psi_R^i\gamma^\mu P_R\psi_R^jZ^{\prime}_\mu\,,
\ee
where $P_{L,R}=(1\mp\gamma_5)/2$. Here $\psi$ represent classes of
fermions with the same electric charge, i.e., $u,\,d,\,e,\,\nu$, while $i,j$
are generation indices: $1,2,3$. For instance $u_3=t$ and $e_2=\mu$.
The couplings $\Delta^{ij}_{L,R}(Z^{\prime})$ are given as follows
\be\label{Zprimecouplings0}
\Delta^{ij}_L(Z^{\prime})=g_{Z^\prime} z_{\psi_L^i}\delta^{ij},\qquad \Delta^{ij}_R(Z^{\prime})=g_{Z^\prime}z_{\psi_R^i}\delta^{ij}\,,
\ee
with $z_{\psi_L^i}$ and $z_{\psi_R^i}$ being fermion charges under $\text{U(1)}^\prime$.

It should be noted that $Z^{\prime}$ couplings are flavour
conserving at this stage, but the charges  $z_{\psi_L^i}$ and  $z_{\psi_R^i}$ will generally depend on the generation index $i$.
This is necessary in order to generate FCNCs mediated by the $Z^\prime$ through the rotation of fermions to the mass eigenstate basis. We will perform this
rotation in Section~\ref{ROTMASS}, but as the cancellation of anomalies
can be performed in the flavour basis, we continue our discussion in this basis.

{ In the present paper we will consider only scenarios with SM fermions, except
possibly RH neutrinos. We leave for the future the analysis of scenarios with additional fermions which are vectorial with respect to the SM gauge group but not with respect to $\text{U(1)}^\prime$ so that possible left-over gauge anomalies can be cancelled by these new fermions. }

As the $Z^\prime$ is a singlet under the SM gauge group, the LH  leptons in a given doublet $\ell_i$ with $i=1,2,3$ must have the same $\text{U(1)}^\prime$ charges and similar for the members of LH quark doublets $q_i$. On the other hand, the RH leptons $\nu_i$ and $e_i$ can have different $\text{U(1)}^\prime$ charges and the same for the RH quarks $u_i$ and $d_i$. We will allow all these charges to be generation dependent. The
RH neutrinos will sometimes turn out to be relevant for the cancellation of the gauge anomalies.

Denoting by $q_i$ and $\ell_i$ LH doublets and by $u_i$, $d_i$, $\nu_i$
and $e_i$ RH singlets allows us to drop the subscripts $L$ and $R$
on the fields that were present in the general formula (\ref{ZprimeL}).

Concerning the normalization of the hypercharge $Y$,  we  use
\be
Q=T_3+Y,
\ee
so that for SM fermions we have independently of the generation index $i$
\be\label{eq:SMY}
y_{q_i}=1/6, \qquad y_{u_i}=2/3,\qquad y_{d_i}=-1/3,\qquad y_{\ell_i}=-1/2,\qquad y_{e_i} =-1\,.
\ee
\subsection{General Formulae}
There are six gauge anomalies generated by the presence of a $Z^\prime$ \cite{Carena:2004xs}. Four
of them are linear in $\text{U(1)}^\prime$ charges, one is quadratic and one
cubic. It is useful to begin with the first four equations as they
are simpler and teach us already something.

The structure of the four linear equations can be simplified by defining
\be\label{charge1}
z_q=z_{q_1}+z_{q_2}+z_{q_3},\qquad z_u=z_{u_1}+z_{u_2}+z_{u_3},\qquad z_d=z_{d_1}+z_{d_2}+z_{d_3},
\ee
\be\label{charge2}
z_l=z_{l_1}+z_{l_2}+z_{l_3},\qquad z_\nu=z_{\nu_1}+z_{\nu_2}+z_{\nu_3},\qquad z_e=z_{e_1}+z_{e_2}+z_{e_3}.
\ee

The $[SU(3)_C]^2\text{U(1)}^\prime$,  $[SU(2)]^2\text{U(1)}^\prime$ and  $[\text{U(1)}_Y]^2\text{U(1)}^\prime$
anomaly cancellation conditions
can then be written respectively  as
\be
A_{33z}=2z_q-z_u-z_d=0,\label{A33z}
\ee
\be
A_{22z}=3z_q+z_l=0\label{A22z},
\ee
\be\label{A11z}
A_{11z}=\frac{1}{6}z_q-\frac{4}{3}z_u-\frac{1}{3}z_d+\frac{1}{2}z_l-z_e=0.
\ee

The fourth linear condition involves two gravitons and a $Z^\prime$ and is given
by
\be
A_{GGz}=3[2z_q-z_u-z_d]+2z_l-z_e-z_\nu=0\label{AGGza},
\ee
but using (\ref{A33z}) it reduces to
\be
A_{GGz}= 2z_l-z_e-z_\nu=0\label{AGGz}.
\ee

In order to write the remaining two anomaly cancellation conditions in a transparent form we define
\be
z^{(3)}_f=\sum_{i=1,2,3} z^3_{f_i},\qquad z^{(2)}_f=\sum_{i=1,2,3} z^2_{f_i}.
\ee
In terms of this notation,  the $\text{U(1)}_Y [\text{U(1)}^\prime]^2$ anomaly cancellation condition is given by
\be
A_{1zz}=[z^{(2)}_q -  2 z^{(2)}_u + z^{(2)}_d]- [z^{(2)}_l - z^{(2)}_e]=0\label{A1zz},
\ee
and the one for the $[\text{U(1)}^\prime]^3$ anomaly by
\be
A_{zzz}=3[2z^{(3)}_q -  z^{(3)}_u - z^{(3)}_d]+ [2z^{(3)}_l -  z^{(3)}_\nu - z^{(3)}_e]=0\label{Azzz}.
\ee
\subsection{Solution to linear equations}
There are four linear equations with six unknowns. This implies that we can express four of the charges in (\ref{charge1}) and (\ref{charge2}) in terms of
 remaining ones. We consider two cases.

{\bf d-case:}

We choose as free parameters $z_q$ and $z_d$, then
\be\label{R1}
z_u=2z_q-z_d, \qquad z_l=-3 z_q, \qquad z_e=z_d-4 z_q,\qquad z_\nu=-2 z_q-z_d\,.
\ee

{\bf u-case:}

We choose as free parameters $z_q$ and $z_u$, then
\be\label{R2}
z_d=2z_q-z_u, \qquad z_l=-3 z_q, \qquad z_e=-z_u-2 z_q,\qquad z_\nu=-4 z_q+z_u\,.
\ee

We emphasize that the simple relations (\ref{R1}) and (\ref{R2}) are model-independent. Moreover, it turns out that the solutions in the $d$- and $u$-case also solve, in the case of flavour universality, the quadratic and cubic equations (\ref{A1zz})-(\ref{Azzz}).
Furthermore the SM case for the hypercharges can be retrieved from the above relations. This is not surprising, since the same equations as \eq{A33z}, \eq{A22z}, \eq{AGGz} and\eq{Azzz} are obtained in the case of the $\text{U(1)}_Y$ gauge boson, by simply replacing $z_f\to y_f$.
Indeed, the anomaly equations are solved  by choosing $z_{q_i}=1/6$ independently of $i$, as well as $z_{d_i}=-1/3$ in the $d$-case and $z_{u_i}=2/3$ in the $u$-case.

In what follows we will present two simple examples for the $\text{U(1)}^\prime$
charges that satisfy all gauge anomaly cancellation conditions. Afterwards we  introduce the model on which our study is based.
\subsection{Example 1}
This example involves  two parameters
\be
a\equiv z_{q_1},\qquad \text{and} \qquad b\equiv z_{d_1}\,,
\ee
and assumes the universality of $\text{U(1)}^\prime$ charges of the remaining fermions
of a given electric charge as well as vanishing $\text{U(1)}^\prime$ charges of RH neutrinos.  The cancellation of all gauge-anomalies implies then
\be
z_{q_2}=z_{q_3}= -\frac{b}{2},\qquad z_{u_1}=2a-b,\qquad z_{u_2}=z_{u_3}=a-\frac{3}{2}b,\qquad  z_{d_2}=z_{d_3}=\frac{b}{2}-a,
\ee

\be
z_{l_1}=z_{l_2}=z_{l_3}= b-a,\qquad z_{e_1}=z_{e_2}=z_{e_3}=2 (b-a), \qquad
z_{\nu_1}=z_{\nu_2}=z_{\nu_3}=0\,.
\ee

The virtue of this solution is the absence of any new particles. But the
complete flavour universality in the lepton sector and the equality of
charges in the second and third quark generation is its disadvantage.
In particular, the $B$-physics anomalies cannot be addressed. This solution
can then only be vital if the NP effects would dominantly  be
found in the $K$ meson system. We hope this will not be the case and consequently we will not analyze phenomenological consequences of this solution below.

\subsection{Example 2}
Also this example involves two parameters
\be
a\equiv z_{q_1},\qquad \text{and} \qquad b\equiv z_{u_1}\,,
\ee
and assumes the universality of $\text{U(1)}^\prime$ charges of the remaining fermions
of a given electric charge as well as vanishing $\text{U(1)}^\prime$ charges of RH neutrinos.  The cancellation of all gauge-anomalies implies then
\be
z_{q_2}=z_{q_3}= -a+\frac{b}{2},\qquad z_{u_2}=z_{u_3}=-2a+\frac{3}{2}b,\qquad z_{d_1}=2a-b,\qquad  z_{d_2}=z_{d_3}=-\frac{b}{2},
\ee

\be
z_{l_1}=z_{l_2}=z_{l_3}= a-b,\qquad z_{e_1}=z_{e_2}=z_{e_3}=2 (a-b), \qquad
z_{\nu_1}=z_{\nu_2}=z_{\nu_3}=0\,.
\ee
The criticism to the Solution 1 applies also here and we will not consider
it any further, but these two examples demonstrate transparently that
gauge anomaly cancellation implies correlations between the $\text{U(1)}^\prime$
charges of quarks and leptons.

\section{A simple model for gauge anomaly cancellation}\label{sec:3}
\subsection{Cancellation of gauge anomalies}
In order to obtain phenomenologically more interesting scenarios we break flavour
universality in all three families. But in order not to end up with many free
parameters we break the flavour universality in a rather special manner.

To this end let us denote by $f$ one of the following fermions: $f=q,\,u,\,d,\,\ell,\,e,\,\nu$ and let us add a generation index: $f_i$ for $i=1,2,3$. We know that the SM hypercharge is generation universal so that $y_{f_i}=y_f$ $\forall i=1,2,3$.
We write the $z$-charge as:
\be
z_{f_i}=y_f+\epsilon_i
\ee
 so that we have
\be
z_f=\sum_{i=1}^3 z_{f_i}=\sum_{i=1}^3 (y_f+\epsilon_i) = 3 y_f +\sum_{i=1}^3\epsilon_i=3y_f +\epsilon \,,\qquad \epsilon=\sum_{i=1}^3 \epsilon_i\,.
\label{zf}
\ee

It should be noted that the breakdown of flavour universality has been
made in a very special manner. Indeed, the parameters $\epsilon_i$ while generation dependent, are universal within
a given generation. For instance $\epsilon_1$ is the same not only for the members of the LH doublets  $q_1$, $\ell_1$ but also for RH
singlets $u_1$, $d_1$, $\nu_1$ and $e_1$
This implies that the shifts $\epsilon_i$ are vector-like which allows for
a straightforward solution of all anomaly equations. But as hypercharges
of LH and RH fields, as seen in  (\ref{eq:SMY}), differ from
 each other the flavour conserving $Z^\prime$ couplings are not vector-like.

Let us first check that the linear anomaly equations are satisfied keeping in mind that they are satisfied when $z_{f_i}=y_{f_i}=y_f$.
\begin{itemize}
\item
$A_{33z}=2z_q-z_u-z_d=0$.
\\
Substituting $z_f$ from (\ref{zf}) we find
\be
A_{33z}=2(3y_q+\epsilon)-(3y_u+\epsilon)-(3y_d+\epsilon)=3(2y_q-y_u-y_d) =0\,,
\ee
where the last equation holds since the hypercharges solve the anomaly equations.
\item
$A_{22z}=3z_q+z_\ell=0$.
\\
We have
\be
A_{22z}=3(3y_q+\epsilon)+(3y_\ell+\epsilon)=3(3y_q+y_\ell)+4\epsilon=4\epsilon\,,
\ee
where $(3y_q+y_\ell)=0$ again because hypercharges solve the anomaly equations so that this equation is satisfied provided that
$\epsilon=0$.
\item
$A_{11z}=\frac{1}{6}z_q-\frac{4}{3}z_u-\frac{1}{3}z_d+\frac{1}{2}z_\ell-z_e=0$.
\\
We can directly simplify the contribution of the hypercharges using the usual argument and we are left with
\be
A_{11z}=\frac{1}{6}\epsilon-\frac{4}{3}\epsilon-\frac{1}{3}\epsilon+\frac{1}{2}\epsilon-\epsilon=-2 \epsilon
\ee
that is again satisfied if
$\epsilon=0$.
\item $A_{GGz}=2z_\ell-z_e-z_\nu=0$.
\\
The usual reasoning implies that this condition is automatically satisfied.
\end{itemize}
We now turn to the two non linear equations introducing the notation $\epsilon^{(2)}=\sum\limits_{i=1}^3  \epsilon_i ^2$ and $\epsilon^{(3)}=\sum\limits_{i=1}^3  \epsilon_i ^3$.
Let us point out  the following relations:
\bea
z_f^{(2)}&=&\sum_{i=1}^3 z_{f_i}^2=\sum_{i=1}^3 (y_f+\epsilon_i)^2=\sum_{i=1}^3 y_f^2+2y_f \,\sum_{i=1}^3  \epsilon_i +\sum_{i=1}^3  \epsilon_i ^2=3y_f^2+2y_f \epsilon +\epsilon^{(2)}, \label{e2}
\\
z_f^{(3)}&=&\sum_{i=1}^3 z_{f_i}^3=\sum_{i=1}^3 (y_f+\epsilon_i)^3=\sum_{i=1}^3 y_f^3+3y_f \,\sum_{i=1}^3  \epsilon_i^2+3y_f^2 \,\sum_{i=1}^3  \epsilon_i +\sum_{i=1}^3  \epsilon_i ^3 \nn \\
&=&3y_f^3+3y_f\epsilon^{(2)}+3y_f^2\epsilon+\epsilon^{(3)}\,,
\label{e3}
\eea
 and consider the l.h.s of (\ref{A1zz}). Using (\ref{e2}) and considering that the  $\epsilon$-independent terms amount to the contribution of the hypercharge and therefore give zero, we have
\bea
A_{1zz}&=&(2y_q\epsilon+\epsilon^{(2)})-2(2y_u\epsilon+\epsilon^{(2)})+(2y_d\epsilon+\epsilon^{(2)})-[(2y_\ell\epsilon+\epsilon^{(2)})
-(2y_\nu \epsilon+\epsilon^{(2)})] \nn \\
&=&\epsilon \big( [2y_q-4y_u+2y_d]-2[y_\ell-y_e]\big)+\epsilon^{(2)}(1-2+1-1+1)=0\,.
\eea
Indeed,  the first term vanishes because  from the linear equations
$\epsilon=0$, and the second one is zero because the coefficients sum to zero.

Finally, let us consider $A_{zzz}=3[2z_q^{(3)}-z_u^{(3)}-z_d^{(3)}]+[2z_\ell^{(3)}-z_\nu^{(3)}-z_e^{(3)}]$.
Using (\ref{e3}), cancelling the  $\epsilon$-independent terms and those proportional to $\epsilon$ we are left with
\bea
A_{zzz}&=& 3\Big[2 (3y_q\epsilon^{(2)}+\epsilon^{(3)})-(3y_u\epsilon^{(2)}+\epsilon^{(3)})-(3y_d\epsilon^{(2)}+\epsilon^{(3)})\Big] \nn \\&+&
2(3y_\ell\epsilon^{(2)}+\epsilon^{(3)})-(3y_e\epsilon^{(2)}+\epsilon^{(3)})-(3y_\nu\epsilon^{(2)}+\epsilon^{(3)})
 \\
&=& 3\epsilon^{(2)}\Big[3(2y_q-y_u-y_d)+2y_\ell-y_e-y_\nu\Big] +\epsilon^{(3)}[3(2-1-1)+2-1-1]=0\,. \nn
\eea
This holds because the coefficients of $\epsilon^{(2)}$ and $\epsilon^{(3)}$
vanish, as one can verify using (\ref{eq:SMY}) in the  case of the coefficient
of   $\epsilon^{(2)}$   and evidently
for the coefficient of  $\epsilon^{(3)}$.

Until now we worked in the flavour basis but for phenomenology we have
  to rotate fermions to the mass eigenstate basis.


\subsection{Rotation of fermions to the mass eigenstate basis}\label{ROTMASS}
\subsubsection{General formulation}
We denote fermion mass matrices in the flavour basis by $\hat M_\psi$ treating
 neutrinos as Dirac particles. Diagonalizing them by unitarity matrices
$V_L^\psi$ and $V_R^\psi$ as follows
\be
(\hat M_\psi)_D=(V_L^\psi)^\dagger \hat M_\psi V_R^\psi\,,
\ee
we find the familiar CKM \cite{Cabibbo:1963yz,Kobayashi:1973fv} and PMNS \cite{Pontecorvo:1957cp,Maki:1962mu} matrices
\be\label{CKMPMNS}
 V_{\rm CKM}= (V_L^u)^\dagger V_L^d,\qquad  U_{\rm PMNS}= (V_L^e)^\dagger V_L^\nu\,.
\ee
Note the difference in the common definition of the PMNS matrix relative to the
CKM matrix, so that in fact $U^\dagger_{\rm PMNS}$ corresponds to the CKM matrix.

Next, in the presence of flavour non-universal $Z^\prime$ couplings,
flavour-violating $Z^\prime$ couplings are generated in the fermion mass eigenstate basis so that instead of (\ref{Zprimecouplings0}) we have now  \cite{Langacker:2000ju}
\be\label{Zprimecouplings}
\Delta^{ij}_L(Z^{\prime})=g_{Z^\prime} B^{ij}_L(\psi_L),\qquad \Delta^{ij}_R(Z^{\prime})=g_{Z^\prime} B^{ij}_R(\psi_R)\,,
\ee
where
\be\label{Bij}
 B^{ij}_L(\psi_L)=[(V_L^\psi)^\dagger \hat Z^{\psi}_L  V_L^\psi]^{ij},\qquad
 B^{ij}_R(\psi_R)=[(V_R^\psi)^\dagger  \hat Z^{\psi}_R  V_R^\psi]^{ij}.
\ee
$\hat Z_{\psi,L}$ and $\hat Z_{\psi,R}$ are diagonal matrices
\bea
 {\hat Z_L^{\psi}} =\text{diag}[z_{\psi_L^1},\,z_{\psi_L^2},\,z_{\psi_L^3}] \,,\nn \\
  {\hat Z_R^{\psi}} =\text{diag}[z_{\psi_R^1},\,z_{\psi_R^2},\,z_{\psi_R^3}]\,, \eea
with the diagonal elements composed of the $\text{U(1)}^\prime$ charges of the SM fermions.
Let us then specify these equations to the simple model in question.
\subsubsection{Rotation to mass eigenstates: quarks}\label{sec:quarkrot}
Let us consider the Yukawa Lagrangian for the SM fermions in the presence of a single Higgs doublet
 $\phi$ and denote with
$\displaystyle{\left(
\begin{array}{c} U \\ D  \end{array}
\right)_L}$ a generic LH doublet. We have two
possibilities to build up a quantity invariant under $SU(2)$
transformations, i.e.:
\be
\phi^\dagger \left(
\begin{array}{c} U \\ D  \end{array}
\right)_L\,, \hskip 1cm \phi^T \tilde\epsilon \left(
\begin{array}{c} U \\ D  \end{array}
\right)_L \,,\ee
where T means the transpose and $\tilde\epsilon$ is the
totally antisymmetric tensor:
\be \tilde\epsilon=\left(
\begin{array}{cc} 0 & 1 \\ -1 & 0  \end{array}
\right) \label{epsilon} \,.
\ee

Considering explicitly the case of quarks and leptons, the Yukawa Lagrangian  reads:
\be {\cal L}_{yuk}=\sum_{i,j=1}^3 \Big\{ -\left(C_e \right)_{ij} {\bar
e}_{i\, R} \phi^\dagger \left(
\begin{array}{c} \nu_{e_j L} \\ e_{j \,L}   \end{array}
\right)+\left(C_q^\prime\right)_{ij} {\bar
u}_{i\, R} \phi^T\tilde \epsilon \left(
\begin{array}{c} u_{j L} \\ d_{j \,L}^\prime    \end{array}
\right) -\left(C_q\right)_{ij} {\bar
d}_{i\, R}^\prime \phi^\dagger \left(
\begin{array}{c} u_{j L} \\ d_{j \,L}^\prime   \end{array}
\right)
+ \text{h.c.} \Big\}\label{yuk} \ee

We next assume that for up-quarks flavour and mass eigenstates are equal
to each other. In the SM this assumption would be immaterial because
of the flavour universality of the $Z^0$ couplings. But for the $Z^\prime$ interactions it is a model assumption that in the basis in which the up-quark mass matrix is diagonal $Z^\prime$ couplings to up-quarks are flavour conserving. This is the simplest  assumption in order to generate flavour violation mediated by a $Z^\prime$ in the down-quark sector which is necessary if we want to address $B$ physics anomalies, possible  $\epe$ anomaly and decays like $\kpn$ and
$\klpn$.

For the time being we  indicate  by a prime that the down-type quarks are at this stage still
in the flavour basis before they will be rotated to mass eigenstates.
 ${\cal L}_{yuk}$ is invariant under $\text{U(1)}^\prime$ if the $z$-charges  satisfy for each generation ($i=1,2,3$) the requirements:
\bea
z_H=z_{q_i}-z_{d_i} \,,\nn \\
z_H=z_{u_i}-z_{q_i} \,,\nn \\
z_H=z_{\ell_i}-z_{e_i}\,, \label{zHiggs}
\eea
where $z_H$ denotes the Higgs $\text{U(1)}^\prime$-charge.
Recalling that in our model the fermion $z$-charges are given by $z_{f_i}=y_f+\epsilon_i$,
if we fix $z_H=y_H$, i.e. the Higgs SM hypercharge, we see that the relations (\ref{zHiggs}) are automatically satisfied because the Yukawa Lagrangian is invariant under the SM gauge group $\text{U(1)}_Y$.
For example, to demonstrate this explicitly, we consider the first generation.
Then:
$z_{d_1}=y_d+\epsilon_1$, $z_{q_1}=y_q+\epsilon_1$ so that  the first equation in (\ref{zHiggs}) is $z_{q_1}-z_{d_1}=y_q+\epsilon_1-y_d-\epsilon_1=y_q-y_d=y_H=z_H$.

In order to find the interactions of the $Z^{\prime}$ in the quark mass eigenstate basis  we assume, as stated above, $V^u_L=\hat 1$ and $V^u_R=\hat 1$ and concentrate our discussion on the down quarks that are rotated as follows
\be
\left(
\begin{array}{c}  d^\prime \\ s^\prime \\ b^\prime \end{array}
\right)_L =V_{\rm CKM} \left(
\begin{array}{c}  d \\ s \\ b \end{array}
\right)_L \,, \qquad
\left(
\begin{array}{c}  d^\prime \\ s^\prime \\ b^\prime \end{array}
\right)_R =V^d_{R} \left(
\begin{array}{c}  d \\ s \\ b \end{array}
\right)_R \,.
 \label{CKM}
\ee
In writing the first equation we used the fact that in the up-basis\footnote{See for example \cite{Aebischer:2017ugx,Aebischer:2015fzz} for a discussion on basis definitions of Wilson coefficients.} we have $V^u_L=\hat 1$ and consequently $V_L^d=V_{\rm CKM}$ as seen in (\ref{CKMPMNS}).
$V^d_{R}$ is a unitary matrix which although present in the SM, does
not appear in the SM interactions because of the absence of RH charged
currents in the SM and flavour universality of RH $Z^0$
couplings.

Recalling that
\be
 {\hat Z_L^{d}} =\text{diag}[z_{q_1},\,z_{q_2},\,z_{q_3}] \,,\qquad
  {\hat Z_R^{d}} =\text{diag}[z_{d_1},\,z_{d_2},\,z_{d_3}]\,, \ee
we find for $i\not=j$
\be\label{Zprime couplings3}
\Delta_L^{ij}(Z^{\prime},d_L)= g_{Z^\prime} \sum_{\alpha=1}^3 \epsilon_\alpha [\lambda_{u_\alpha}^{(ij)}]_L,\qquad
\Delta_R^{ij}(Z^{\prime},d_R)=g_{Z^\prime} \sum_{\alpha=1}^3 \epsilon_\alpha [\lambda_{u_\alpha}^{(ij)}]_R,
\ee
where the parameters $[\lambda_{u_\alpha}^{(ij)}]_{L,R}$ denote
\be
[\lambda_{u_\alpha}^{(ij)}]_L=(V_{\rm CKM})^*_{u_{\alpha}i}(V_{\rm CKM})_{u_{\alpha}j}\,,\qquad
[\lambda_{u_\alpha}^{(ij)}]_R=(V^d_R)^*_{u_{\alpha}i}(V^d_R)_{u_{\alpha}j}\,.
\ee
It should be noted that the terms in the $z_i$-charges involving hypercharges
disappeared because of their flavour universality. Furthermore, the parameters $\epsilon_\alpha$ are the same in the LH and RH couplings,
a specific property of our model. We recall that the cancellation of gauge anomalies requires their sum to vanish so that

\be
\boxed{\epsilon_3=-\epsilon_1-\epsilon_2,}
\label{rel-epsilon}
\ee
\newline
with $\epsilon_{1,2}$ being real rational numbers.

Adopting the standard CKM phase convention, where the 5 relative phases of the quark fields are adjusted to remove 5 complex phases from the CKM matrix, we have no more
freedom to remove the 6 complex phases from  $V_R^d$.  In the standard CKM basis  $V_R^d$  can then be parametrised as follows \cite{Buras:2010pz,Blanke:2011ry}
\be
V^d_R = D_U V_R^0 D^\dagger_D\,,
\ee
where $V_R^0$ is a ``CKM-like'' mixing matrix, containing only three real mixing angles and one non-trivial phase.
The diagonal matrices $D_{U,D}$ contain the remaining CP-violating phases. Choosing the standard parametrisation for $V_R^0$ we have
\be
V_R^0 =
\left(
\begin{array}{ccc}
\tc_{12}\tc_{13}&\ts_{12}\tc_{13}&\ts_{13}e^{-i\phi}\\ -\ts_{12}\tc_{23}
-\tc_{12}\ts_{23}\ts_{13}e^{i\phi} &\tc_{12}\tc_{23}-\ts_{12}\ts_{23}\ts_{13}e^{i\phi}&
\ts_{23}\tc_{13}\\
\ts_{12}\ts_{23}-\tc_{12}\tc_{23}\ts_{13}e^{i\phi}&-\ts_{23}\tc_{12} -\ts_{12}\tc_{23}\ts_{13}e^{i\phi}&\tc_{23}\tc_{13}
\end{array}
\right)\,,
\label{eq:Vtrgen}
\ee
and
\be
D_U={\rm diag}(1, e^{i\phi^u_2}, e^{i\phi^u_3})\,,  \qquad
D_D={\rm diag}(e^{i\phi^d_1}, e^{i\phi^d_2}, e^{i\phi^d_3})\,.
\label{eq:Dphases}
\ee

 Inserting these formulae into (\ref{Zprime couplings3}) results in
 flavour violating $Z^{\prime}$ couplings to down-quarks.
The  flavour-conserving ones are modified by the rotations and
can be obtained from (\ref{Zprimecouplings}).
 But the expressions for
them are not transparent. In order to improve on this we
will assume in what follows that the $\tilde s_{ij}$  in the matrix  (\ref{eq:Vtrgen}) are very small  and adopt the simplifying approximation of setting to 1 the cosines and
 retaining only terms linear in $\tilde s_{ij}$.

\noindent
Exploiting then CKM unitarity and the relation (\ref{rel-epsilon}) we
find the following formulae for the $Z^\prime$ couplings that will govern our
numerical analysis in Section~\ref{sec:5}.
\newline

{\bf LH couplings: $i=d,s,b$}
 \bea
 \Delta_L^{ij}(Z^{\prime})&=&g_{Z^\prime} [(2\epsilon_1+\epsilon_2)V_{uj}V^*_{ui}+(\epsilon_1+2\epsilon_2)V_{cj}V^*_{ci}], \qquad   i\neq j
  \\\label{LHQD}
 \Delta_L^{ii}(Z^{\prime})&=&g_{Z^\prime} [\frac{1}{6}+(2\epsilon_1+\epsilon_2)|V_{ui}|^2+(\epsilon_1+2\epsilon_2)|V_{ci}|^2-\epsilon_1-\epsilon_2]\,,
 \eea

{\bf RH couplings: $i=d,s,b$}
\bea
\Delta_R^{sd}(Z^{\prime})&=&g_{Z^\prime} \,e^{i\phi_K} \tilde s_{12}\,(\epsilon_1-\epsilon_2), \label{ZprimesdR}\\
\Delta_R^{bd}(Z^{\prime})&=&g_{Z^\prime} \,e^{i\phi_d} \tilde s_{13}\,(2\epsilon_1+\epsilon_2), \\
\Delta_R^{bs}(Z^{\prime})&=&g_{Z^\prime} \,e^{i\phi_s} \tilde s_{23}\,(\epsilon_1+2\epsilon_2), \\
 \Delta_R^{ii}(Z^{\prime})&=&g_{Z^\prime} [-\frac{1}{3}+ \epsilon_i],
 \eea
where
\bea
\phi_K&=&-\phi_{1d}+\phi_{2d},  \\
\phi_d&=& \phi-\phi_{1d}+\phi_{3d},  \\
\phi_s&=& -\phi_{2d}+\phi_{3d}.
\eea

We note
that the parameters $(\tilde s_{12},\,\phi_K)$ enter only in the Kaon sector,
$(\tilde s_{13},\,\phi_d)$  only in the $B_d$ sector, and $(\tilde s_{23},\,\phi_s)$ only in the $B_s$ sector. This division will make our numerical analysis
more transparent than working directly with (\ref{eq:Vtrgen}).

Finally, for up-quarks we have
\be
 \Delta_L^{ii}(Z^{\prime})= g_{Z^\prime} [\frac{1}{6}+\epsilon_i],\qquad  \Delta_R^{ii}(Z^{\prime})= g_{Z^\prime} [\frac{2}{3}+\epsilon_i],\qquad (i=u,c)\,.
\ee

\subsubsection{Rotation to mass eigenstates: leptons}
Denoting $U_{\rm PMNS}=U$,
rotation to the mass eigenstates in the lepton sector is  usually done through
\begin{align}\label{equ:PMNSdef2}
  \left(\begin{array}{c}
				\nu_{e}\\
				\nu_{\mu}\\
				\nu_{\tau}
				\end{array}
\right)=\left(\begin{array}{ccc}
							U_{e1}&U_{e2}&U_{e3}\\
							U_{\mu 1}&U_{\mu 2}&U_{\mu 3}\\
							U_{\tau 1}&U_{\tau 2}&U_{\tau 3}
							\end{array}
							\right)\left(\begin{array}{c}
				\nu_{1}\\
				\nu_{2}\\
				\nu_{3}
				\end{array}
\right)\,,
\end{align}
\noindent
where $\nu_\ell$ with $\ell = e,\mu,\tau$ are flavour eigenstates and $\nu_i$, with $i = 1,2,3$ mass eigenstates. Comparing with the CKM matrix in
(\ref{CKM}) we note that whereas the PMNS matrix relates neutrinos
in the mass and interaction (flavour) bases, the CKM matrix is doing it for
down-quarks. This is only the convention used in the literature which assumes
that $V_L^e=\hat 1$ and has no impact on physics implications within the SM.

Here we prefer to consider scenarios in which $V_L^\nu=\hat 1$, so that, the rotation
is made on charged leptons instead of neutrinos. In this case we have
\be
U_{\rm PMNS}= (V_L^e)^\dagger\,,
\ee
and consequently instead of (\ref{equ:PMNSdef2})
\begin{align}\label{equ:PMNSdef3}
  \left(\begin{array}{c}
				e^\prime\\
				\mu^\prime\\
				\tau^\prime
				\end{array}
\right)=\left(\begin{array}{ccc}
					U^*_{e1}&U^*_{\mu 1}&U^*_{\tau 1}\\
					U^*_{e2}&U^*_{\mu 2}&U^*_{\tau 2}\\
					U^*_{e3}&U^*_{\mu 3}&U^*_{\tau 3}
							\end{array}
							\right)\left(\begin{array}{c}
				e\\
				\mu\\
				\tau
				\end{array}
\right)\,.
\end{align}
\noindent
$e^\prime,\mu^\prime,\tau^\prime$ are flavour eigenstates and $e,\mu,\tau$  mass eigenstates.

We find then
for $i\not=j$
\be\label{Zprime couplings3L}
\Delta_L^{ij}(Z^{\prime},e_L)=g_{Z^\prime} \sum_{\alpha=1}^3 \epsilon_\alpha [\lambda_{\nu_\alpha}^{(ij)}]_L,\qquad
\Delta_R^{ij}(Z^{\prime},e_R)=g_{Z^\prime} \sum_{\alpha=1}^3 \epsilon_\alpha [\lambda_{\nu_\alpha}^{(ij)}]_R,
\ee
where the parameters $[\lambda_{\nu_\alpha}^{(ij)}]_{L,R}$ denote:
\be
[\lambda_{\nu_\alpha}^{(ij)}]_L=(U^\dagger_{\rm PMNS})_{i\nu_{\alpha}}(U^\dagger_{\rm PMNS})^*_{j\nu_{\alpha}}\,,\qquad [\lambda_{\nu_\alpha}^{(ij)}]_R=(V^e_R)^*_{\nu_{\alpha}i}(V^e_R)_{\nu_{\alpha}j}\,.
\ee
The matrix $V^e_R$ is analogous to $V^d_R$ and can be parametrized as
the latter one with three new mixing angles and six complex phases.

The LH couplings to charged leptons are then given by:
\newline

{\bf LH couplings: $i=e,\mu,\tau$}
 \bea\label{LHQD1}
 \Delta_L^{ij}(Z^{\prime})&=&g_{Z^\prime}\,[(2\epsilon_1+\epsilon_2)U_{i 1}U^*_{j1}+(\epsilon_1+2\epsilon_2)U_{i2}U^*_{j2}], \qquad   i\neq j
  \\\label{LHQD2}
  \Delta_L^{ii}(Z^{\prime})&=&g_{Z^\prime}\,[-\frac{1}{2}+(2\epsilon_1+\epsilon_2)|U_{i1}|^2+(\epsilon_1+2\epsilon_2)|U_{i2}|^2-\epsilon_1-\epsilon_2]\,.
 \eea

Similarly,

{\bf RH couplings: $i=e,\mu,\tau$}
\bea
\Delta_R^{\mu e}(Z^{\prime})&=&g_{Z^\prime}\,e^{i\phi_{12}} \tilde t_{12}\,(\epsilon_1-\epsilon_2), \\
\Delta_R^{\tau e}(Z^{\prime})&=&g_{Z^\prime}\,e^{i\phi_{13}} \tilde t_{13}\,(2\epsilon_1+\epsilon_2), \\
\Delta_R^{\tau\mu}(Z^{\prime})&=&g_{Z^\prime}\,e^{i\phi_{23}} \tilde t_{23}\,(\epsilon_1+2\epsilon_2), \\
 \Delta_R^{ii}(Z^{\prime})&=&g_{Z^\prime}\,[-1 + \epsilon_i],
 \eea
where $\phi_{ij}$ and $\tilde t_{ij}$ replace $\phi_{K,d,s}$ and $\tilde s_{ij}$ of the
quark sector. Thus
$(\tilde t_{12},\,\phi_{12})$ parametrize $\mu\to e$ transitions,
$(\tilde t_{13},\,\phi_{13})$ describe $\tau\to e$ transitions
 and $(\tilde t_{23},\,\phi_{23})$ enter  $\tau\to \mu$ ones.

Finally, for neutrinos,
\be\label{neutrinos}
 \Delta_L^{\nu_i\nu_i}(Z^{\prime})= g_{Z^\prime}\,[-\frac{1}{2}+\epsilon_i],\qquad (i=\nu_1,\nu_2, \nu_3)\,.
\ee
There are no light RH neutrinos and we will assume that the RH ones are so heavy that they cannot {contribute} to the processes considered by us.

In order to evaluate the couplings in (\ref{LHQD1}) and (\ref{LHQD2})
we will use central values of the relevant entries and parameters in the PMNS matrix as resulting from the fit in \cite{Esteban:2018azc}
\be
\sin^2\theta_{12}=0.310,\qquad \sin^2\theta_{23}=0.582, \qquad  \sin^2\theta_{13}=0.0224, \qquad  \delta=217^\circ,
\ee
\be
|U_{e 1}|=0.821, \qquad |U_{e 2}|=0.550, \qquad |U_{e 3}|=0.150,
\ee
\be
|U_{\mu 1}|=0.290, \qquad |U_{\mu 2}|=0.590, \qquad |U_{\mu 3}|=0.754,
\ee
\be
|U_{\tau 1}|=0.491, \qquad |U_{\tau 2}|=0.592, \qquad |U_{\tau 3}|=0.639.
\ee
The values given above satisfy the PMNS unitarity constraint within a few percent which is much less than uncertainties in separate entries.

\section{Considered observables}\label{sec:4}
\subsection{Preliminaries}
 We refrain, with the exception of lepton flavour violation, from listing the formulae for
 observables entering our analysis as they can be found in the same notation in \cite{Buras:2012jb,Buras:2012dp,Buras:2015jaq}. For $\epe$ we use the results obtained in \cite{Aebischer:2018csl,Aebischer:2018quc,Aebischer:2018rrz}. We match the ${Z^\prime}$ model directly onto the Weak effective theory (WET) at the electroweak scale. The Wilson coefficients of the WET are then evolved to the corresponding scales via the one-loop QCD anomalous dimensions \cite{Aebischer:2017gaw,Jenkins:2017dyc}. In the following we report the lepton flavour violating observables used in the numerical analysis.
\boldmath
\subsection{$\mu\to e \gamma$, $\tau\to \mu \gamma$ and $\tau\to e \gamma$}
\unboldmath
We will use the formulae of  \cite{Lindner:2016bgg}. We find in the case
of the $\mu\to e\gamma$ decay
\be
\mathcal{B}(\mu\to e \gamma)=\frac{3(4\pi)^3\alpha}{4G^2_F}
\left[|A^M_{e\mu}|^2+|A^E_{e\mu}|^2\right],
\ee
where
\be
A^M_{e\mu}= -\frac{1}{96\pi^2M^2_{Z^\prime}}\sum_{f=e,\mu,\tau}
\left[\Delta^{fe*}_V(Z^\prime)\Delta^{f\mu}_V(Z^\prime)\left(1-3\frac{m_f}{m_\mu}\right)+\Delta^{fe*}_A(Z^\prime)\Delta^{f\mu}_A(Z^\prime)\left(1+3\frac{m_f}{m_\mu}\right)\right],
\ee
\be
A^E_{e\mu}= \frac{i}{96\pi^2M^2_{Z^\prime}}\sum_{f=e,\mu,\tau}
\left[\Delta^{fe*}_A(Z^\prime)\Delta^{f\mu}_V(Z^\prime)\left(1-3\frac{m_f}{m_\mu}\right)+\Delta^{fe*}_V(Z^\prime)\Delta^{f\mu}_A(Z^\prime)\left(1+3\frac{m_f}{m_\mu}\right)\right].
\ee
The couplings $\Delta^{ij}_V$ and  $\Delta^{ij}_A$ are defined as
\be\label{DVA}
 \Delta_V^{ij}(Z')= \Delta_R^{ij}(Z')+\Delta_L^{ij}(Z'),\qquad
\Delta_A^{ij}(Z')= \Delta_R^{ij}(Z')-\Delta_L^{ij}(Z').
\ee
The summation is over internal charged leptons in the loop. This result is also valid for new heavy charged
leptons provided $m_f/M_{Z^\prime}\le 0.2 $. For heavier leptons exact expressions in \cite{Lindner:2016bgg} have to be used. For $\tau\to \mu \gamma$ and $\tau\to e \gamma$ obvious changes of flavour indices have to be made.

The present upper bounds read:\cite{TheMEG:2016wtm}
\be
\mathcal{B}( \mu \to e \gamma )\le 4.2\times 10^{-13}, \label{boundmueg}
\ee
and \cite{Aubert:2009ag}
\be
\mathcal{B}( \tau \to \mu \gamma ) \le 4.4 \times 10^{-8},\qquad
\mathcal{B}( \tau \to e \gamma ) \le 3.3 \times 10^{-8}\,.
\ee

\boldmath
\subsection{Three-body Lepton decays}
\unboldmath
For a decay  $\ell_j\to \ell_i \bar \ell_l \ell_k  $
we have the following contributing operators and their Wilson coefficients with $X,Y=L,R$
\be
[Q_{VXY}]^{ij}_{kl}=(\bar\psi_i\gamma^\mu P_X \psi_j)(\bar\psi_k\gamma_\mu P_Y \psi_l)\,,\qquad [C_{VXY}]^{ij}_{kl}=\frac{\Delta^{ij}_X(Z^{\prime})\Delta^{kl}_Y(Z^{\prime})}{M^2_{Z^\prime}}.
\ee

Calculating the branching ratios for the most interesting decays with
$k=l$ we have to take into account the presence of two identical leptons in the
final state that requires the introduction of a factor $1/2$ at the level of the branching ratio. Moreover always
two diagrams, differing by the interchange of identical leptons, contribute. They interfere with each other for $VLL$ and $VRR$ cases but not for $VLR$ and $VRL$. We find then
\be \label{DecayA}
\mathcal{B}(\tau^-\to\mu^-\mu^+\mu^-)=
\frac{m_\tau^5}{1536\pi^3 \Gamma_{\tau}} \left[2 |C_{VLL}|^2
+ 2 |C_{VRR}|^2 + |C_{VLR}|^2 + |C_{VRL}|^2\right]^{\mu\tau}_{\mu\mu}\,,
\ee
with analogous expressions for $\mu^-\to e^-e^+e^-$ and $\tau^-\to e^-e^+e^-$.
$\Gamma_\tau$ is the total $\tau$ decay width.
This formula agrees with the one in \cite{Straub:2018kue,Brignole:2004ah},
and  can also be derived from the expressions in
\cite{Crivellin:2013hpa} with the factors of 2, exhibited here,  hidden in the Wilson coefficients.

Present bounds are \cite{Amhis:2019ckw}:
\bea
\mathcal{B}(\tau^-\to\mu^-\mu^+\mu^-)&<& 1.2 \cdot 10^{-8}\,, \label{boundtau3mu} \\
\mathcal{B}(\tau^-\to e^-e^+e^-)&<& 1.4 \cdot 10^{-8}\,, \label{boundtau3e} \\
\mathcal{B}(\mu^-\to e^-e^+e^-)&<& 1.0 \cdot 10^{-12}\,, \label{boundmu3e} \,\,\eea
that update the more conservative PDG bounds  \cite{Tanabashi:2018oca}.
\boldmath
\subsection{${\mu-e}$ conversion in nuclei}
\unboldmath
We give next the formula for the $\mu$-$e$ conversion in nuclei, that is
\be
\mu +(A,Z) \rightarrow e + (A,Z)\,,
\ee
where $Z$
and $A$ denote the proton and atomic numbers in a nucleus,
respectively.
Adapting general formulae in \cite{Hisano:1995cp} and keeping the dominant tree-level contributions we find
\begin{eqnarray}\label{eq:Gmue}
\Gamma(\mu\rightarrow e)
&=& \frac{\alpha^3}{16\pi^2}\frac{Z_{eff}^4}{Z}|F(q)|^2~\frac{m_{\mu}^5}{M^4_{Z^\prime}}
\left [|\Delta_L^{\mu e}(Z^\prime)|^2+|\Delta_R^{\mu e}(Z^\prime)|^2\right]
\nonumber \\
&&\times
|(2Z+N)\Delta_V^{uu}(Z^\prime)+(Z+2N)\Delta_V^{dd}(Z^\prime)|^2
,
\end{eqnarray}
\noindent
with $\Delta^{ij}_V$ given in (\ref{DVA}).
$Z$ and $N$ denote the proton and neutron numbers in a nucleus,
respectively. $Z_{eff}$ is  an effective parameter and $F(q^2)$
 the nuclear form factor.
The branching ratio used by experimentalists is defined by
 \begin{equation}
   \mathcal{B}(\mu\to e) = \frac{\Gamma(\mu\rightarrow e)}{\Gamma_{\rm capt}}\,,
 \end{equation}
 where $\Gamma_{\rm capt}$ is  the muon capture rate.
 For the case of $^{48}_{22}{\rm{Ti}}$,  one finds $Z_{eff}=17.6$ and
$F(q^2\simeq -m_{\mu}^2) \simeq 0.54$ \cite{Bernabeu:1993ta} and $\Gamma_{\rm capt}\simeq(2.590 \pm 0.012) \cdot 10^6/s$ \cite{PhysRevC.35.2212}. The present experimental bound for the branching ratio reads \cite{Bertl:2006up}
 \be
 \mathcal{B}(\mu\to e)\le \mathcal{O}(10^{-12})\,.
 \ee

\boldmath
\subsection{$(g-2)_\mu$ and $(g-2)_e$}
\unboldmath
Using general formulae in \cite{Lindner:2016bgg} we find for $a_\mu=(g-2)_\mu/2$
\be
\Delta a_\mu(Z^\prime)\approx -\frac{1}{16\pi^2}\frac{m_\mu^2}{M^2_{Z^\prime}}
\sum_{f=e,\mu,\tau}
\left[\left|\Delta^{f\mu}_V(Z^\prime)\right|^2\left(\frac{2}{3}-\frac{m_f}{m_\mu}\right)+\left|\Delta^{f\mu}_A(Z^\prime)\right|^2\left(\frac{2}{3}+\frac{m_f}{m_\mu}\right)\right]
\ee
with $\Delta^{ij}_V$ and  $\Delta^{ij}_A$ given in (\ref{DVA}). The summation is over internal charged leptons in the loop. This result is also valid for new heavy charged
leptons provided $m_f/M_{Z^\prime}\le 0.2 $. For heavier leptons exact expressions in \cite{Lindner:2016bgg} have to be used. For $(g-2)_e$ one just replaces $\mu$ by $e$.

\section{Various coupling scenarios}\label{sec:5}
\subsection{Preliminaries and general strategy for  numerics}
In Section~\ref{sec:3}
we have derived the couplings of the $Z^\prime$ to
 SM quarks and leptons as well as to RH neutrinos assuming that they
are Dirac particles.

It should be
 recalled that the parameters $\epsilon_i$  are the same in the quark and lepton sectors which in our simple model is a direct consequence of the cancellation of gauge anomalies. This then implies that
we should expect correlations between various observables not only separately
 within the lepton and quark systems but in particular between lepton and quark
 observables.

Before entering numerics it is strategically useful  to count the full number
of free parameters and subsequently define a few simple scenarios in which
some of these parameters vanish. In this manner the number of correlations between various observables is increased.

Let us then count the number of free parameters:
\begin{itemize}
\item
4 real parameters entering both quark and lepton couplings
\be\label{four}
g_{Z^\prime},\qquad M_{Z^\prime},\qquad \epsilon_1,\qquad \epsilon_2\,.
\ee
Even if all formulae listed above depend only on the ratio $g_{Z^\prime}/M_{Z^\prime}$, the renormalization group effects depend only on  $M_{Z^\prime}$ and these are
two independent parameters.
\item
1 complex phase in the PMNS matrix in LH couplings. The remaining
parameters in the latter couplings are already measured parameters
of the CKM and PMNS matrices.
\item
3 mixing angles and three complex phases in RH quark couplings:
\be\label{quarks}
 (\tilde s_{12},\,\phi_K), \qquad  (\tilde s_{13},\,\phi_d),\qquad
(\tilde s_{23},\,\phi_s).
\ee
\item
3 mixing angles and three  complex phases in RH lepton couplings:
\be\label{leptons}
 (\tilde t_{12},\,\phi_{12}), \qquad  (\tilde t_{13},\,\phi_{13}),\qquad
(\tilde t_{23},\,\phi_{23}).
\ee
\end{itemize}

We observe then that in full generality
we have 10 real parameters and 6 phases to our disposal. This could
appear as very many. Yet, by considering $K$, $B_d$ and $B_s$ meson
systems with $\Delta F=1$ and $\Delta F=2$ transitions and the charged lepton flavour violating decays,  $\mu-e$ conversion as well as $(g-2)_{e,\mu}$, there is  a sufficient number
of observables so that not only the parameters in question can be bounded
but also correlations between various observables can be predicted.

In what follows, we will consider four constrained scenarios.
{Except for the last one, we will adopt a common strategy for numerics.
We will perform a simplified analysis of $\Delta F=2$ observables. The relevant formulae for these when a new $Z^\prime$ gauge boson is present can be found in Section 3.3 of \cite{Buras:2012jb} to which we address the reader.}
We consider  in the Kaon sector $\varepsilon_K$ and  in the $B_{d,s}$ sectors the neutral meson mass differences
$\Delta M_{d,s}$, as well, as the CP asymmetries  $S_{\psi K_S}$
and $S_{\psi\phi}$ in order to identify oases in the space of new parameters discussed in
the previous section for which these five observables are consistent with experiment.
 To this end, we set all other input parameters at their central values.
But to take partially hadronic
and experimental uncertainties into account we require the theory
to reproduce the data for $\varepsilon_K$ within $\pm 10\%$, $\Delta M_{s,d}$ within $\pm 5\%$ and the
data on $S_{\psi K_S}$ and $S_{\psi\phi}$ within experimental
$2\sigma$. We choose a larger uncertainty for  $\varepsilon_K$  than for $\Delta M_{s,d}$ because of its strong $\vcb^4$ dependence. For the neutral Kaon mass difference $\Delta M_K$ we will
only require the agreement within $\pm 25\%$ because of potential long
distance uncertainties.

Specifically, our search is governed by the following allowed ranges:
\be\label{C1}
16.0/{\rm ps}\le \Delta M_s\le 19.5/{\rm ps},
\quad  0.01\le S_{\psi\phi}\le 0.10,
\ee
\be\label{C2}
0.455/{\rm ps}\le \Delta M_d\le 0.556/{\rm ps},\quad
0.665\le S_{\psi K_S}\le 0.733 .
\ee

\be\label{C3}
0.75\le \frac{\Delta M_K}{(\Delta M_K)_{\rm SM}}\le 1.25,\qquad
2.0\times 10^{-3}\le |\varepsilon_K|\le 2.5 \times 10^{-3}.
\ee
Moreover, in the Kaon sector, we impose that the short-distance contribution to the ${\cal  B}(K_L \to \mu^+ \mu^-)$ satisfies the bound \cite{Isidori:2003ts}:
\be
{\cal  B}(K_L \to \mu^+ \mu^-)_{SD}<2.5 \cdot 10^{-9} \,\,.
\ee
In what follows,  we will first identify the
region in the space of the parameters $\epsilon_1$ and $\epsilon_2$
allowed by the constraints in (\ref{C1})-(\ref{C3}) for each scenario. Subsequently, we
will investigate correlations between various observables in this region.

For the numerical analysis of the model of Section~\ref{sec:3} we consider a relatively light $Z'$ setting $M_{Z^\prime}=1$ TeV and we vary the gauge coupling in the range $g_{Z^\prime}\in[0.01,\,1]$. This will
allow us to neglect RG effects from Yukawa and electroweak interactions that could also imply correlations between observables, thereby exhibiting
the correlations implied by the cancellation of gauge anomalies. Therefore,
as already stated above, the matching is performed directly
onto the WET below the EW scale and only QCD RG effects are taken into account.
We summarize our intput in Tables~\ref{tab:numinput1} and \ref{tab:numinput2}.

\begin{table}[h]
\centering
{\small
\begin{tabular}{|l|l|l|}
\hline
$G_F = 1.16637(1)\times 10^{-5}\gev^{-2}$\hfill 	&  $M_Z = 91.188(2) \gev$\hfill
& $M_W = 80.385(15) \gev$\\
$\sin^2\theta_W = 0.23116(13)$\hfill & $\alpha(M_Z) = 1/127.94$\hfill & $\alpha_s(M_Z)= 0.1184(7) $\\
\hline
$m_e=0.511\mev$\hfill & $m_\mu=105.66\mev$\hfill & $m_\tau=1776.9(1)\mev$\\
$m_u(2\gev)=2.16(11)\mev $ \hfill & $m_c(m_c) = 1.279(13) \gev$ \hfill
& $m_t(m_t) = 163(1)\gev$\\
$m_d(2\gev)=4.68(15)\mev$\hfill & $m_s(2\gev)=93.8(24) \mev$\hfill &
$m_b(m_b)=4.19^{+0.18}_{-0.06}\gev$\\
\hline
$m_{K^\pm}=493.68(2)\mev$\hfill & $m_{K^0}=497.61(1)\mev$\hfill &
\\
$m_{B_d}=5279.62(15)\mev$\hfill &
$m_{B_s} = 5366.82(22)\mev$\hfill &
\\
\hline
$\Delta M_K= 0.5292(9)\times 10^{-2} \,\text{ps}^{-1}$\hfill &
 $\Delta M_d = 0.5055(20) \,\text{ps}^{-1}$\hfill &
$\Delta M_s = 17.757(21) \,\text{ps}^{-1}$ \\
$|\eps_K|= 2.228(11)\times 10^{-3}$\hfill & $S_{\psi K_S}= 0.699(17)$\hfill &
$S_{\psi\phi}= 0.054(20)$\\
$\sin^2\left(\theta_{12}\right)= 0.307\pm 0.013$\hfill & $\sin^2\left(\theta_{23}\right)= 0.506\pm 0.04$\hfill & $\sin^2\left(\theta_{13}\right)= 0.0212\pm 0.0008$\hfill
\\
\hline
\end{tabular}}
\caption{Values of theoretical quantities used for the numerical analysis.\label{tab:numinput1}}
\end{table}

\begin{table}[t]
\centering
\renewcommand{\arraystretch}{1.3}
\begin{tabular}{|l|l|l|}
\hline
$F_{B_d}$ = $190.5(13)\mev$ \hfill &
$F_{B_s}$ = $230.7(12)\mev$ \hfill &
$F_K = 156.1(11)\mev$\hfill\\
$\hat B_{B_d} =1.232(53)$ \hfill &  $\hat B_{B_s} =1.222(61)$\hfill
& $\hat B_K= 0.766(10)$ \\
$F_{B_d} \sqrt{\hat B_{B_d}} = 210.6(55)\mev$\hfill &
$F_{B_s} \sqrt{\hat B_{B_s}} = 256.1(57)\mev$\hfill &
$\xi = 1.21(2)$\\
$\eta_{cc}=1.87(76)$\hfill & $\eta_{ct}= 0.496(47)$\hfill &
$\eta_{tt}=0.5765(65)$\\
$\eta_B=0.55(1)$\hfill & $\phi_{\epsilon}=43.51(5)^\circ$\hfill & $\kappa_\epsilon=0.94(2)$ \\
$|V_{us}|=0.2254(4)$\hfill & $|V^\text{nom}_{ub}|=3.7\times10^{-3}$\hfill
&$|V^\text{nom}_{cb}|=42.0\times10^{-3}$\\
\hline
\end{tabular}
\caption{Numerical values used for the numerical analysis.\label{tab:numinput2}}
\end{table}

\subsection{Different Scenarios}
We will consider a number of different scenarios that we want to describe
briefly here.
\subsubsection{Scenario A}
In this scenario flavour violation is present only in the {\em LH} couplings. This is achieved by setting all parameters $\tilde s_{ij}$ and  $\tilde t_{ij}$
to zero such that all flavour-violating RH couplings vanish.
This leaves us with only four real parameters in (\ref{four}) and
the PMNS phase.

It should be noted that
the flavour conserving  RH couplings are generally non-zero. In fact
inspecting the formulae for these couplings, it is clear that it is impossible to set them all to zero.

There are two important implications of this structure:
\begin{itemize}
\item
The absence of RH flavour-violating couplings implies the absence of left-right
operators contributing to $\Delta F=2$ processes. As the hadronic matrix
elements of these operators are, in particular in the case of the
$K^0-\bar K^0$ system, strongly enhanced, their absence allows to satisfy
constraints from $\varepsilon_K$ and the mass differences $\Delta M_K$,
$\Delta M_d$ and $\Delta M_s$ easier than in the subsequent scenarios.
\item
Non-vanishing flavour conserving RH currents together with flavour-violating LH currents imply the presence of the dominant QCD $(Q_6)$ and electroweak penguin $(Q_8)$ operators contributing to $\epe$ thereby
allowing to address the $\epe$-anomaly.
\item
Moreover the small number of free parameters implies a number of correlations.
\end{itemize}

\subsubsection{Scenarios B1 and B2}
In these scenarios
flavour violation is present
in both LH and RH quark currents, but for leptons only in
LH currents. At first sight six additional parameters in (\ref{quarks})
enter the phenomenology. However,
the presence of left-right operators implies that it is very difficult to
satisfy the constraints in the $K^0-\bar K^0$ mixing for $M_{Z^\prime}=\ord (1\tev)$,  so that, the $\Delta_R^{sd}$ coupling has to be strongly suppressed or even
eliminated. This can be done by setting

\be
\hspace{-4cm}\bullet\hspace{4cm} \tilde s_{12}=0,\qquad  (\text{Scenario B1})
\ee

or

\be
\hspace{-4cm}\bullet\hspace{4cm}\epsilon_1=\epsilon_2, \qquad (\text{Scenario B2}).
\ee
In both scenarios only four additional parameters relative to Scenario A are
present. Moreover, in Scenario B2 the number of free parameters is reduced
through the relation $\epsilon_1=\epsilon_2$.

\subsubsection{Scenario C}
In this  scenario, with both LH and RH currents, it is possible, as explained in the Appendix A,  to eliminate  $\Delta F=2$ constraints in Kaon and $B_{s,d}$ systems which allows to obtain larger NP effects in $\Delta F=1$ transitions.
This allows to see the correlations between various observables in different
systems even better than in the previous scenarios.

\section{Numerical Analysis}\label{sec:5a}
\subsection{Results in Scenario A}
\begin{figure}[hb]
\begin{center}
\includegraphics[width = 0.58\textwidth]{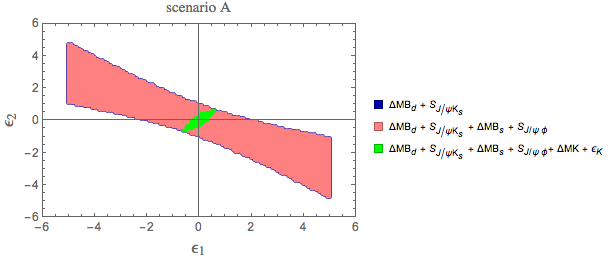}
\includegraphics[width = 0.36\textwidth]{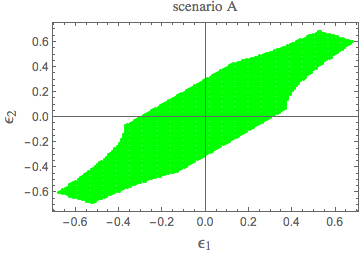}
\caption{\baselineskip 10pt  \small  Scenario A. Left panel: Allowed region in the $(\epsilon_1,\,\epsilon_2)$ plane, after imposing $\Delta F=2$ constraints, as indicated in the legend. Right panel: Zoom of the green zone in the left panel. }\label{fig:epspar}
\end{center}
\end{figure}

\begin{figure}[ht]
\begin{center}
\includegraphics[width = 0.4\textwidth]{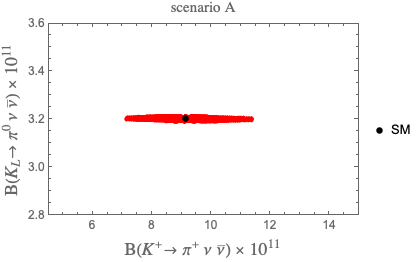}
\caption{\baselineskip 10pt  \small  Scenario A. Correlation between ${\cal B}(K_L \to \pi^0 \nu {\bar \nu})$ and ${\cal B}(K^+ \to \pi^+ \nu \bar \nu)$. The black dot represents the central value of the SM prediction.}\label{fig:KplusvsKL}
\end{center}
\end{figure}

The impact of  the constraints (\ref{C1})-(\ref{C3}) on the parameter space in this scenario can be deduced from Fig.~\ref{fig:epspar}.
In the left plot, we show the allowed ranges in the $(\epsilon_1,\epsilon_2)$ space resulting separately from  $\Delta F=2$ constraints in the $B_d,\,B_s, K$ systems, respectively. The relations (\ref{C1}) and (\ref{C2}) produce the  blue and pink overlapping regions, while the constraints in (\ref{C3}) significantly restrict such a space to the green zone, that we display enlarged in the right panel.
The impact on the observables related to the $B_d$ and $B_s$ system is small. For example, we find that
effects on ${\cal B}(B_d \to \mu^+ \mu^-)$ and ${\bar {\cal B}}(B_s \to \mu^+ \mu^-)$ are at most $2\%$ in this scenario, while effects on the Wilson coefficients $C_9$ and $C_{10}$ entering in the description of  rare $b \to s$ decays are negligible. For this reason, we do not show the corresponding plots.
As for Kaon decays, while effects on $K_L \to \pi^0 \nu {\bar \nu}$ are tiny,
significant deviations from the SM at the level of $20\%$  are still allowed for $K^+ \to \pi^+ \nu \bar \nu$ by $\Delta F=2$ constraints, as can be observed from Fig.~\ref{fig:KplusvsKL}.
This scenario does not predict an enhancement of $|\epe|$ with respect to the SM prediction:
the largest deviation from the SM  is $\simeq 2 \cdot 10^{-5}$, two orders of magnitude below its experimental value.
\begin{figure}[ht]
\begin{center}
\includegraphics[width = 0.4\textwidth]{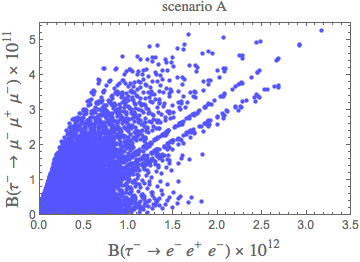} \hskip0.5cm
\includegraphics[width = 0.4\textwidth]{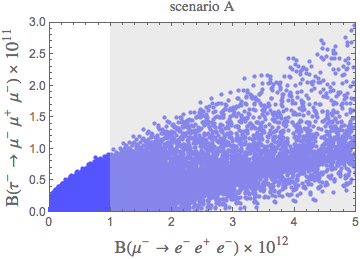}
\caption{\baselineskip 10pt  \small Scenario A. Correlation between ${\cal B}(\tau^-  \to \mu^- \mu^+ \mu^-) $ and ${\cal B}(\tau^-  \to e^-e^+ e^-)$ (left panel) and between ${\cal B}(\tau^-  \to \mu^- \mu^+ \mu^-) $ and ${\cal B}(\mu^-  \to e^-e^+ e^-)$ (right panel). The gray shaded region in the right panel is excluded by the experimental bound ${\cal B}(\mu^-  \to e^-e^+ e^-)<1 \cdot 10^{-12}$.}\label{fig:lepton}
\end{center}
\end{figure}
\begin{figure}[ht]
\begin{center}
\includegraphics[width = 0.4\textwidth]{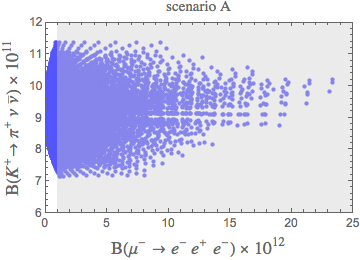}\hskip0.5cm
\includegraphics[width = 0.4\textwidth]{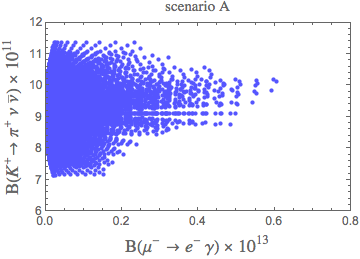}\\
\includegraphics[width = 0.4\textwidth]{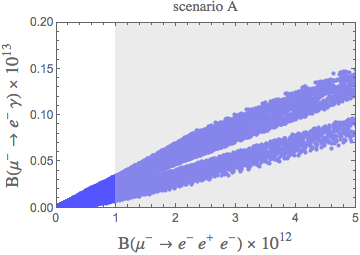}
\caption{\baselineskip 10pt  \small  Scenario A. Correlation between ${\cal B}(K^+ \to \pi^+ \nu \bar \nu) $ and ${\cal B}(\mu^-  \to e^-e^+ e^-)$ (upper left panel) and between ${\cal B}(K^+ \to \pi^+ \nu \bar \nu) $ and ${\cal B}(\mu^-  \to e^-\gamma)$ (upper right panel).  Correlation between ${\cal B}(\mu^-  \to e^-\gamma) $ and ${\cal B}(\mu^-  \to e^-e^+ e^-)$
(lower panel). The gray shaded regions  are excluded by the experimental bound ${\cal B}(\mu^-  \to e^-e^+ e^-)<1 \cdot 10^{-12}$.}\label{fig:Kvslepton}
\end{center}
\end{figure}

Fig.~\ref{fig:lepton} displays the results for flavour violating lepton decays. The left panel shows that $\tau$ decays are predicted well below the experimental upper bounds in (\ref{boundtau3mu})-(\ref{boundtau3e}).
On the other hand, ${\cal B}(\mu^-  \to e^-e^+ e^-)$ shown in the right panel can easily be close to  the experimental bound  (\ref{boundmu3e}) and even exceed it.

Also ${\cal B}(\mu^-  \to e^- \gamma)$ is predicted below the bound in (\ref{boundmueg}), as can be inferred from the upper right panel in Fig.~\ref{fig:Kvslepton}. In this figure, we display the correlations among lepton decays $\mu^-  \to e^-e^+ e^-$, $\mu^-  \to e^- \gamma$ and the mode $K^+ \to \pi^+ \nu {\bar \nu}$. The  upper panels show, that in this scenario,  the experimental bound on both leptonic decays have  no impact on the possibility that ${\cal B}(K^+ \to \pi^+ \nu {\bar \nu})$ could significantly deviate from its SM value. On the other hand, the bound on ${\cal B}(\mu^-  \to e^-e^+ e^-)$
reduces the allowed range for ${\cal B}(\mu^-  \to e^- \gamma)$ (lower panel); however,  this branching ratio is much smaller than the experimental upper bound, as already observed.

\begin{figure}[ht]
\begin{center}
\includegraphics[width = 0.4\textwidth]{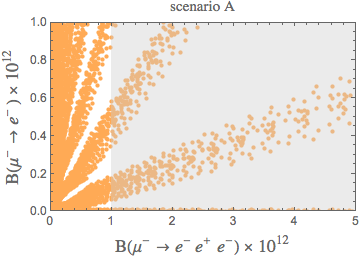}
\hskip0.5cm
\includegraphics[width = 0.4\textwidth]{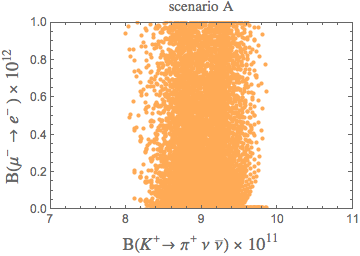}
\caption{\baselineskip 10pt  \small  Scenario A.   Correlation between ${\cal B}(\mu^-  \to \e^-) $  and  ${\cal B}(\mu^-  \to e^-e^+ e^-)$ (left panel). The gray shaded region  is excluded by the experimental bound ${\cal B}(\mu^-  \to e^-e^+ e^-)<1 \cdot 10^{-12}$.
  Correlation between
${\cal B}(\mu^-  \to \e^-) $  and ${\cal B}(K^+ \to \pi^+ \nu \bar \nu) $ (right panel).}\label{fig:mueconvA}
\end{center}
\end{figure}

 Comparing  Fig.~\ref{fig:mueconvA} with Fig.~\ref{fig:KplusvsKL} we observe
  that   the upper bound on $\mu-e$ conversion in the ballpark of $10^{-12}$
  reduces significantly the allowed range for ${\cal B}(K^+ \to \pi^+ \nu \bar \nu)$. On the other hand, as seen in this figure  it has presently no impact
  on the allowed range for  $\mu^-  \to e^-e^+ e^-$. This is also the case
  of $\mu^-  \to e^- \gamma$ not shown here.

\subsection{Results in Scenario B1}
\begin{figure}[ht]
\begin{center}
\includegraphics[width = 0.7\textwidth]{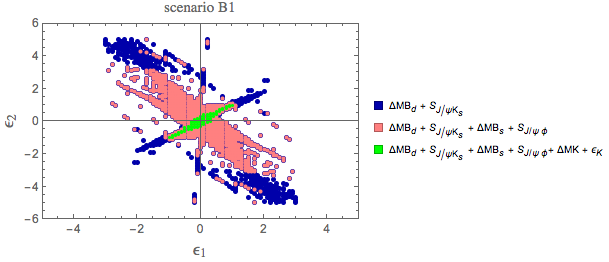}\\
\vskip 0.4cm
\includegraphics[width = 0.7\textwidth]{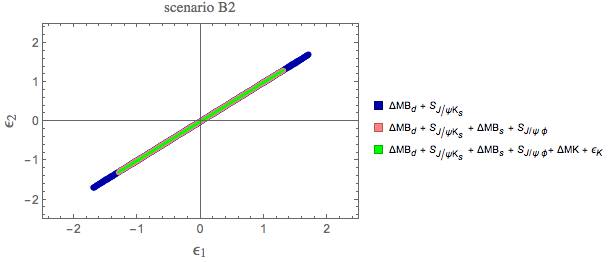}
\caption{\baselineskip 10pt  \small  Scenarios B1 (upper panel) and B2 (lower panel).  Allowed region in the $(\epsilon_1,\,\epsilon_2)$ plane, after imposing $\Delta F=2$ constraints, as indicated in the legend. }\label{fig:epsparB1}
\end{center}
\end{figure}

The result of the subsequent application of the constraints (\ref{C1})-(\ref{C3}) is shown in the upper panel of Fig.~\ref{fig:epsparB1}.
\begin{figure}[ht]
\begin{center}
\includegraphics[width = 0.4\textwidth]{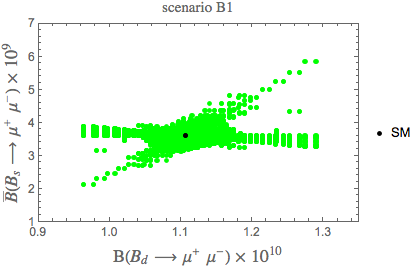}\hskip 0.5cm
\includegraphics[width = 0.4\textwidth]{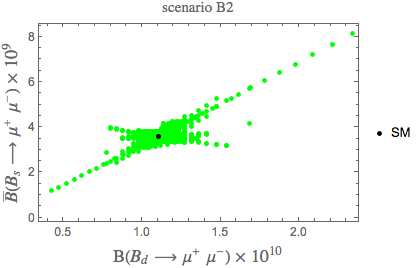}
\caption{\baselineskip 10pt  \small  Scenarios B1 (left panel) and B2 (right panel). Correlation between ${\cal B}(B_d \to \mu^+ \mu^-)$ and ${\cal \bar B}(B_s \to \mu^+ \mu^-)$. The black dot represents the central value of the SM prediction.}\label{fig:BsvsBdB1}
\end{center}
\end{figure}
\begin{figure}[ht]
\begin{center}
\includegraphics[width = .57\textwidth]{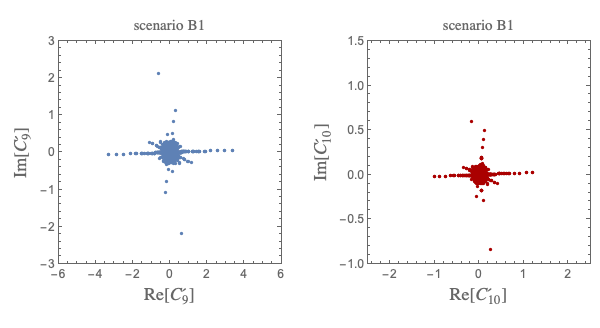}\\
\includegraphics[width = .9\textwidth]{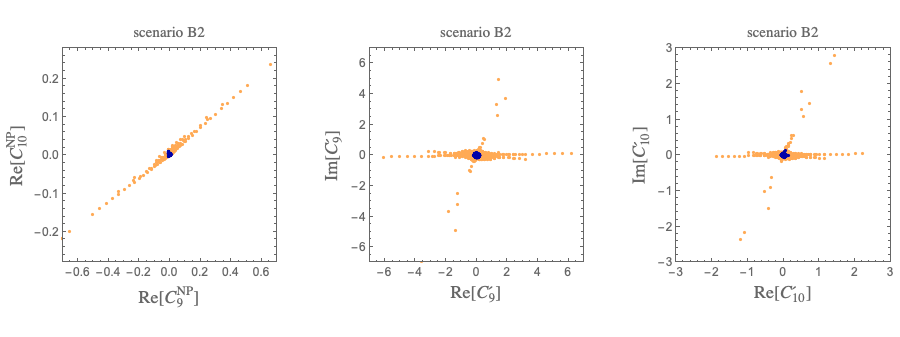}
\caption{\baselineskip 10pt  \small Scenarios B1 (upper plots) and B2 (lower plots). Wilson coefficients  that appear in the  $b \to s \ell^+  \ell^-$ effective Hamiltonian. Only those sensibly different from zero are displayed: $C_9^\prime$ and $C_{10}^\prime$ (not present in the SM) in  scenario B1 (upper plots) and in B2 (lower middle and right plots). In the case of B2 deviation from the SM value of ${\rm Re}[C_9]$ and ${\rm Re}[C_{10}]$ are also shown (lower left plot).  }\label{fig:C9C10B1}
\end{center}
\end{figure}
At odds with the previous scenario, in this case one can have significant deviations from the SM predictions in $B_{d,s}$ observables.
This is for instance the case for the rare decays $B_{d,s} \to \mu^+ \mu^-$. As Fig.~\ref{fig:BsvsBdB1} (left panel) shows, the branching ratios of these modes can largely deviate from the SM results whose central values are represented by the black dot in that figure.

The most recent  averages from the data of  CMS, LHCb and ATLAS
 \cite{Chatrchyan:2013bka,CMS:2014xfa,Aaij:2017vad,Aaboud:2018mst}
 have been presented in \cite{Aebischer:2019mlg} with the result
 \be\label{LHCb2}
\overline{\mathcal{B}}(B_{s}\to\mu^+\mu^-) = (2.71\pm 0.40) \times 10^{-9},
\qquad
\mathcal{B}(B_{d}\to\mu^+\mu^-) =(1.01\pm 0.81)\times 10^{-10}.
\ee
In the case of
$B_{s}\to\mu^+\mu^-$,
we observe a departure of the data from the SM prediction by about
$2\sigma$. This favoures NP scenarios  in which the branching ratio for this decay is suppressed relatively to its SM value
\cite{Aebischer:2019mlg}
\be\label{LHCbTH}
\overline{\mathcal{B}}(B_{s}\to\mu^+\mu^-)_{\rm SM} = (3.67\pm 0.15) \times 10^{-9},
\qquad
\mathcal{B}(B_{d}\to\mu^+\mu^-) =(1.14\pm 0.12)\times 10^{-10}.
\ee

The relevant formulae for these observables when a new $Z^\prime$ contribution is added can be found in Section 3 of \cite{Buras:2012jb}, as well as, those of the Wilson coefficients of the $b \to s \ell^+ \ell^-$ effective Hamiltonian that we are going to consider.
Among such coefficients $C_9$ and $C_{10}$ are already present in the SM, and we shall consider only the shift with respect to their SM values, denoted by $C_9^{\rm NP}$ and $C_{10}^{\rm NP}$. $C_9^\prime$ and $C_{10}^\prime$ are instead absent in the SM, so these represent pure NP quantities.
Various analyses have been devoted to fit the Wilson coefficients in order to  explain the observed anomalies in a number of $b \to s$  related observables.  Among the possible solutions, it has been  pointed out that a large negative value of $C_9^{\rm NP}$ or a large $C_9^\prime$ could be effective.
Our results for the various coefficients in scenario B1 are displayed in the upper panel of Fig.~\ref{fig:C9C10B1}. Deviations from the SM are  negligible
  for $C_9^{\rm NP}$ and $C_{10}^{\rm NP}$ but could be relevant in the
  case of  $C_9^\prime$ and $C_{10}^\prime$ and we display only them.

\begin{figure}[ht]
\begin{center}
\includegraphics[width = 0.4\textwidth]{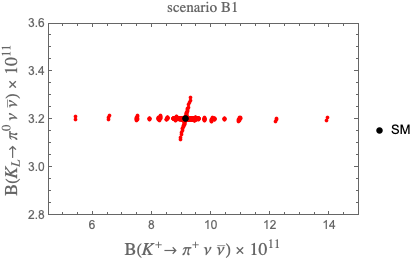}\hskip 0.5 cm \includegraphics[width = 0.4\textwidth]{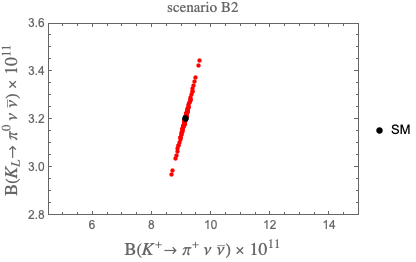}
\caption{\baselineskip 10pt  \small Scenarios B1 (left panel) and B2 (right panel). Correlation between ${\cal B}(K_L \to \pi^0 \nu {\bar \nu})$ and ${\cal B}(K^+ \to \pi^+ \nu \bar \nu)$. The black dot represents the central value of the SM prediction. }\label{fig:KplusvsKLB1}
\end{center}
\end{figure}

In the Kaon sector, we find that ${\cal B}(K_L \to \pi^0 \nu {\bar \nu})$ displays small deviations from the SM, while ${\cal B}(K^+ \to \pi^+ \nu {\bar \nu})$ can be strongly enhanced or suppressed with respect to the SM result,  when ${\cal B}(K_L \to \pi^0 \nu {\bar \nu})$ is SM-like, while it is approximately stable on the SM value when ${\cal B}(K_L \to \pi^0 \nu {\bar \nu})$ deviates from the SM central value. This pattern of correlation is shown in the left panel of Fig.~\ref{fig:KplusvsKLB1}.

\begin{figure}[ht]
\begin{center}
\includegraphics[width = 0.4\textwidth]{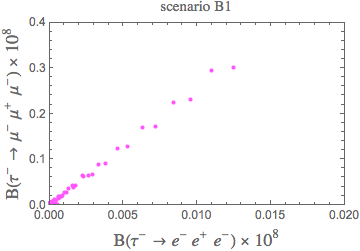} \hskip0.5cm
\includegraphics[width = 0.4\textwidth]{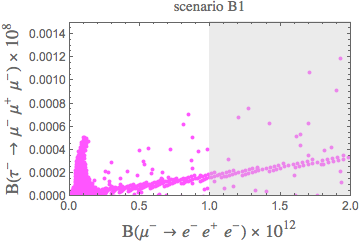}\\
\vskip 0.4cm
\includegraphics[width = 0.4\textwidth]{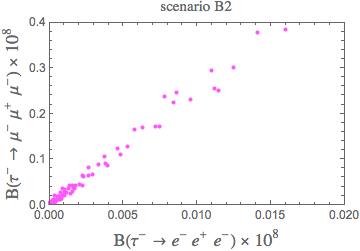} \hskip0.5cm
\includegraphics[width = 0.4\textwidth]{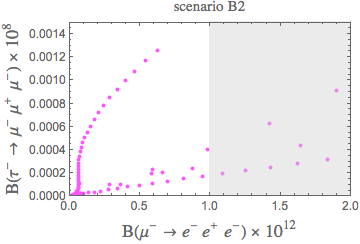}
\caption{\baselineskip 10pt  \small  Scenarios B1 (upper plots) and B2 (lower plots). Correlation between ${\cal B}(\tau^-  \to \mu^- \mu^+ \mu^-) $ and ${\cal B}(\tau^-  \to e^-e^+ e^-)$ (left panel) and between ${\cal B}(\tau^-  \to \mu^- \mu^+ \mu^-) $ and ${\cal B}(\mu^-  \to e^-e^+ e^-)$ (right panel). The gray shaded region in the right panel is excluded by the experimental bound ${\cal B}(\mu^-  \to e^-e^+ e^-)<1 \cdot 10^{-12}$.}\label{fig:leptonB1}
\end{center}
\end{figure}

As for scenario A, also in B1 lepton flavour violating $\tau$ decays are predicted below the experimental upper limit, while ${\cal B}(\mu^- \to e^- e^+ e^-)$ can saturate or even exceed its experimental upper bound. Correlations among these modes are displayed in the upper panel of Fig.~\ref{fig:leptonB1}.

\begin{figure}[ht]
\begin{center}
\includegraphics[width = 0.4\textwidth]{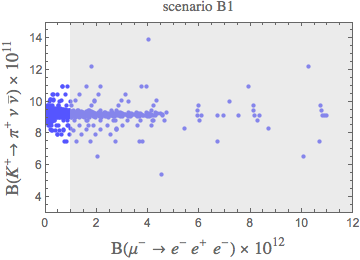}\hskip0.5cm
\includegraphics[width = 0.4\textwidth]{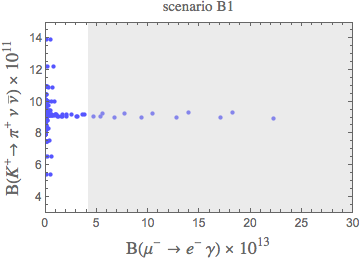}\\
\vskip 0.4cm
\includegraphics[width = 0.4\textwidth]{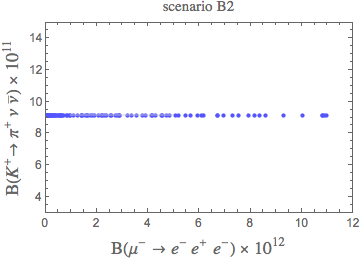}\hskip0.5cm
\includegraphics[width = 0.4\textwidth]{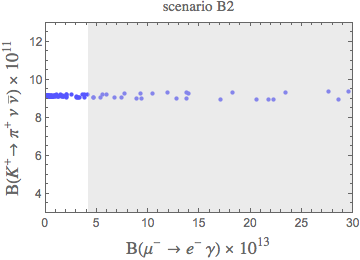}
\caption{\baselineskip 10pt  \small Scenarios B1 (upper plots) and B2 (lower plots). Correlation between ${\cal B}(K^+ \to \pi^+ \nu \bar \nu) $ and ${\cal B}(\mu^-  \to e^-e^+ e^-)$ (left panels) and between ${\cal B}(K^+ \to \pi^+ \nu \bar \nu) $ and ${\cal B}(\mu^-  \to e^-\gamma)$ (right panels).  The gray shaded regions  are excluded by the experimental bounds ${\cal B}(\mu^-  \to e^-e^+ e^-)<1 \cdot 10^{-12}$ and ${\cal B}(\mu^-  \to e^-\gamma)<4.2 \cdot 10^{-13}$. }\label{fig:KvsleptonB1}
\end{center}
\end{figure}

In contrast to the case of scenario A, in B1 ${\cal B}(\mu^-  \to e^-\gamma)$ can saturate or even exceed the experimental upper bound.
Correlations among the lepton observables ${\cal B}(\mu^- \to e^- e^+ e^-)$ or  ${\cal B}(\mu^-  \to e^-\gamma)$ and ${\cal B}(K^+ \to \pi^+ \nu \bar \nu)$ are shown in Fig.~\ref{fig:KvsleptonB1} (upper plots).
 The experimental bound on ${\cal B}(\mu^-  \to e^-\gamma)$  has no impact on the $K^+$ branching ratio. However, one can observe that, the bound on
    ${\cal B}(\mu^-  \to e^-e^+e^-)$  allows only the values for  ${\cal B}(K^+ \to \pi^+ \nu \bar \nu)$ close to its SM value.

\begin{figure}[ht]
\begin{center}
\includegraphics[width = 0.4\textwidth]{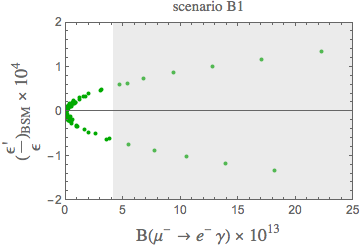}\hskip0.5cm
\includegraphics[width = 0.4\textwidth]{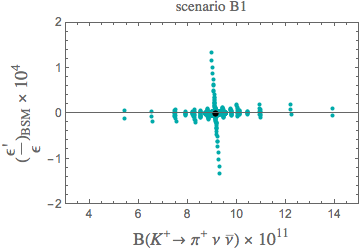}\\
\vskip 0.4cm
\includegraphics[width = 0.4\textwidth]{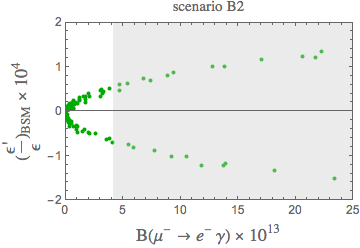}\hskip0.6cm
\includegraphics[width = 0.4\textwidth]{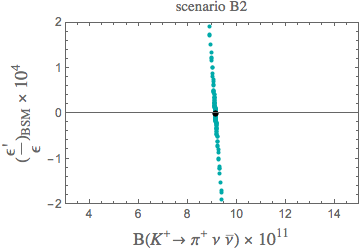}
\caption{\baselineskip 10pt  \small Scenarios B1 (upper plots) and B2 (lower plots). Correlations involving the   $\left( \epe \right)_{\rm BSM}$ (the NP contribution to $\epe$). Correlation with  ${\cal B}(\mu^-  \to e^-\gamma)$ (left panels) and with ${\cal B}(K^+ \to \pi^+ \nu \bar \nu) $ (right panels).
 The gray shaded region  is excluded by the experimental bound ${\cal B}(\mu^-  \to e^-\gamma)<4.2 \cdot 10^{-13}$. }\label{fig:epeB1}
\end{center}
\end{figure}

The situation is different when one considers the ratio $\epe$ and, in particular, its deviation from the SM. As the upper left panel of  Fig.~\ref{fig:epeB1} shows, in B1 $|\left(\epe \right)_{\rm BSM}|$ could reach values as large as $\simeq 1.5 \cdot 10^{-4}$. However, imposing that ${\cal B}(\mu^-  \to e^-\gamma)$ lies below its experimental upper limit, one finds that the maximum deviation in $\epe$ could be $|\left(\epe \right)_{\rm BSM}| \simeq 0.6 \cdot 10^{-4}$. As for the correlation between $\left(\epe \right)_{\rm BSM}$ and ${\cal B}(K^+ \to \pi^+ \nu \bar \nu)$, from the right upper panel of Fig.~\ref{fig:epeB1} one can observe that deviations in one of the two observables with respect to the SM predictions exclude deviations in the other one.

\begin{figure}[ht]
\begin{center}
\includegraphics[width = 0.4\textwidth]{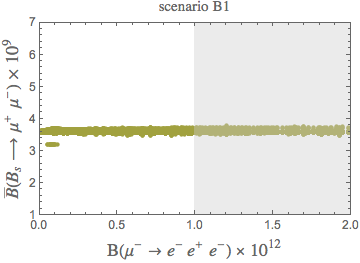}\hskip0.5cm
\includegraphics[width = 0.4\textwidth]{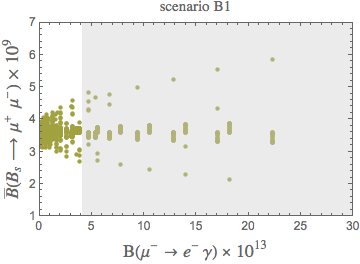}\\
\vskip 0.4cm
\includegraphics[width = 0.4\textwidth]{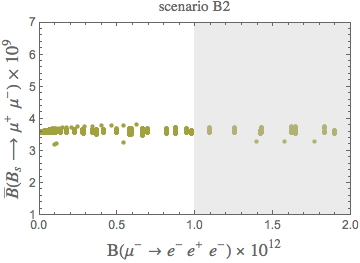}\hskip0.5cm
\includegraphics[width = 0.4\textwidth]{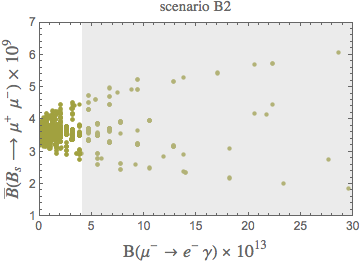}
\caption{\baselineskip 10pt  \small Scenarios B1 (upper plots) and B2 (lower plots). Correlations between  ${\cal \bar B}(B_s \to \mu^+ \mu^-)$ and ${\cal B}(\mu^-  \to e^-e^+ e^-)$ (left panels) and between ${\cal \bar B}(B_s \to \mu^+ \mu^-)$ and ${\cal B}(\mu^-  \to e^-\gamma)$ (right panels).  The gray shaded regions  are excluded by the experimental bounds ${\cal B}(\mu^-  \to e^-e^+ e^-)<1 \cdot 10^{-12}$ and ${\cal B}(\mu^-  \to e^-\gamma)<4.2 \cdot 10^{-13}$. }\label{fig:BsvsleptonsB1}
\end{center}
\end{figure}

 But what is most important is the impact of the bounds on
  leptonic modes on rare $B$ and $K$ decay branching ratios seen already in the  case of Scenario A. First
  one can observe from the upper plots in Fig.~\ref{fig:BsvsleptonsB1} that, even if in B1 ${\cal \bar B}(B_s \to \mu^+ \mu^-)$ can deviate significantly from its SM value, constraints in the lepton sector reduce the size of such a deviation. This is in particular the case when the bound from $\mu^- \to e^- e^+ e^-$ is applied.

\begin{figure}[ht]
\begin{center}
\includegraphics[width = 0.3\textwidth]{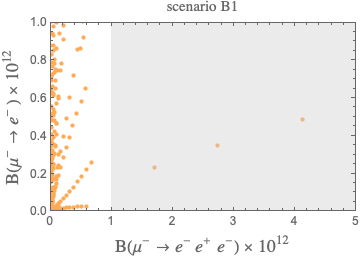}\hskip 0.1cm
\includegraphics[width = 0.3\textwidth]{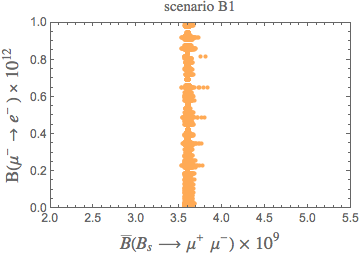} \hskip0.1cm
\includegraphics[width = 0.3\textwidth]{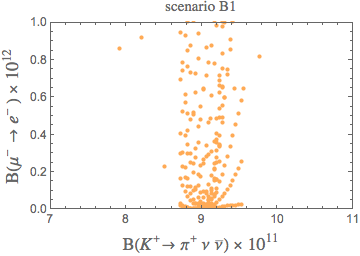}\\
\vskip 0.4cm
\includegraphics[width = 0.3\textwidth]{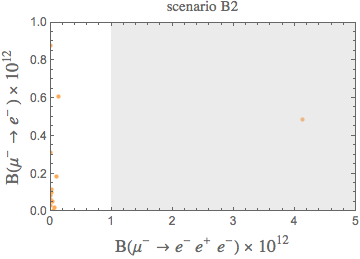}\hskip 0.1cm
\includegraphics[width = 0.3\textwidth]{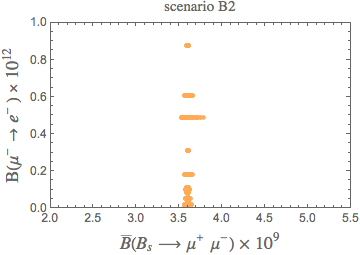} \hskip0.1cm
\includegraphics[width = 0.3\textwidth]{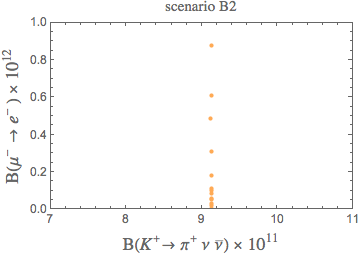}
\caption{\baselineskip 10pt  \small  Scenarios B1 (upper plots) and B2 (lower plots). Correlation between ${\cal B}(\mu^-  \to \e^-) $  and  ${\cal B}(\mu^-  \to e^-e^+ e^-)$ (left panels). The gray shaded region  is excluded by the experimental bound ${\cal B}(\mu^-  \to e^-e^+ e^-)<1 \cdot 10^{-12}$.
Correlations between ${\cal B}(\mu^-  \to \e^-) $  and  ${\cal \bar B}(B_s \to \mu^+ \mu^-)$ (middle panels) and between ${\cal B}(\mu^-  \to \e^-) $  and ${\cal B}(K^+ \to \pi^+ \nu \bar \nu) $ (right panels).}\label{fig:mueconvB1}
\end{center}
\end{figure}

   But even stronger impact comes from $\mu-e$ conversion. Indeed
comparing  Fig.~\ref{fig:mueconvB1} with Figs.~\ref{fig:BsvsBdB1} and \ref{fig:KplusvsKLB1}   we observe
  that   the upper bound on $\mu-e$ conversion in the ballpark of $10^{-12}$
  reduces significantly the allowed ranges for $B_s\to\mu^+\mu^-$ and
  $K^+ \to \pi^+ \nu \bar \nu$ branching ratios. On the other hand, as seen in other plots it has presently no impact
  on the allowed ranges for  $\mu^-  \to e^-e^+ e^-$ and $\mu^-  \to e^- \gamma$.

Scenario B1, similar to Scenario A shows a peculiar feature that often occurs in  our model: in principle large deviations from SM predictions can be found. However, quark and lepton sector act in a complementary way to constrain each other producing observables with values close to the SM ones.


\subsection{Results in Scenario B2}
The parameter space in B2 is simplified because of the assumption $\epsilon_1=\epsilon_2$, as can be observed from the lower panel in Fig.~\ref{fig:epsparB1}.
As in the case of B1, also B2 displays large deviations from the SM in the $B_{d,s}$ sectors.
Fig.~\ref{fig:BsvsBdB1} (right panel) shows that
the rare decays ${\cal B}(B_d \to \mu^+ \mu^-)$ and ${\cal \bar B}(B_s \to \mu^+ \mu^-)$ are approximately linearly correlated, and can strongly deviate from the SM. As for the Wilson coefficients $C_9^{(\prime)},\,C_{10}^{(\prime)}$, in B2 one finds that $\rm{Im}(C_9)\simeq 0$ and $\rm{Im}(C_{10})\simeq 0$,  deviations in $\rm{Re}(C_9)$ and $\rm{Re}(C_{10})$ are of the same size as in B1, while $C_9^\prime,\,C_{10}^\prime$ vary in much larger ranges, as can be seen in Fig.~\ref{fig:C9C10B1} (lower plots).
However, if the experimental bound on ${\cal B}(\mu^-  \to e^-e^+ e^-)$ is imposed, these variations  are practically removed as indicated by the central blue regions in Fig.~\ref{fig:C9C10B1}. This is another example of the mutual role of lepton and quark sectors.

Looking at rare Kaon decays, it can be observed that for $\epsilon_1=\epsilon_2$ the relation between ${\cal B}(K_L \to \pi^0 \nu {\bar \nu})$ and ${\cal B}(K^+ \to \pi^+ \nu \bar \nu)$ is approximately linear. This result is displayed in the right plot in Fig.~\ref{fig:KplusvsKLB1}, which in addition shows that moderate variations with respect to the SM values are possible.

As in B1, among lepton decays the branching fractions that can be large are ${\cal B}(\mu^-  \to e^-e^+ e^-)$, as shown by the lower plots in Fig.~\ref{fig:leptonB1}, and ${\cal B}(\mu^-  \to e^-\gamma)$ (see  lower plots in  Fig.~\ref{fig:KvsleptonB1}). Both reduce the possible range of ${\cal B}(K^+ \to \pi^+ \nu \bar \nu) $, as shown in Fig.~\ref{fig:KvsleptonB1}.
They also reduce the range for $\left(\epe \right)_{\rm BSM}$, the impact of ${\cal B}(\mu^-  \to e^-e^+ e^-)$ being much more constraining than that of ${\cal B}(\mu^-  \to e^-\gamma)$, as shown in the lower left panel of Fig.~\ref{fig:epeB1}. In B2 deviations from the SM are possible in both $\epe$ and ${\cal B}(K^+ \to \pi^+ \nu \bar \nu) $ and the correlation between these two observables is shown in the right lower  panel of Fig.~\ref{fig:epeB1}.

Correlations between lepton decays and ${\cal \bar B}(B_s \to \mu^+ \mu^-)$, shown in Fig.~\ref{fig:BsvsleptonsB1}, follow a pattern similar to that in B1.

\begin{figure}[ht]
\begin{center}
\includegraphics[width = 0.4\textwidth]{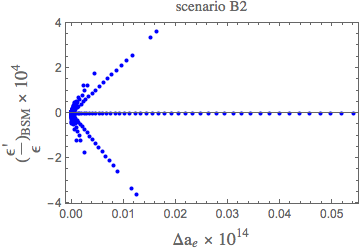}\hskip0.5cm
\includegraphics[width = 0.4\textwidth]{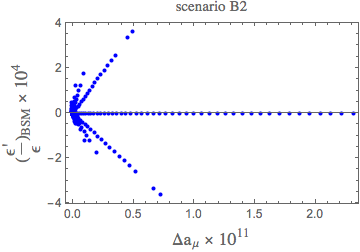}
\caption{\baselineskip 10pt  \small Scenario B2. Correlation between NP contributions to $\epe$ and $a_e$ (left plot) and $a_\mu$ (right plot).}\label{fig:epevsgm2B2}
\end{center}
\end{figure}

 In Fig.~\ref{fig:epevsgm2B2} we show that in this scenario NP effects in $\epe$ can be significant without having any impact on $a_{\mu,e}$. Generally the contributions of NP to $a_{\mu,e}$ are very small in all scenarios considered and this is the only figure displaying them that we show.

 Similar to Scenario B1, as seen in Fig.~\ref{fig:mueconvB1},
 the upper bound on $\mu-e$ conversion
also in this scenario has  a very large impact on $\kpn$ and $B_s\to\mu^+\mu^-$.

\subsection{Results in Scenario C}\label{SC}
\begin{figure}[t]
\begin{center}
\includegraphics[width = 0.47\textwidth]{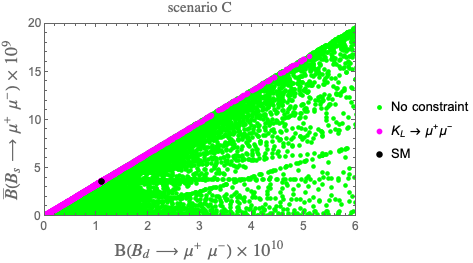}
\caption{\baselineskip 10pt  \small  Scenario C. Correlation between ${\cal B}(B_d \to \mu^+ \mu^-)$ and ${\cal \bar B}(B_s \to \mu^+ \mu^-)$. The black dot represents the central value of the SM prediction. }\label{fig:BsvsBdBB4}
\end{center}
\end{figure}

\begin{figure}[t]
\begin{center}
\includegraphics[width = 0.47\textwidth]{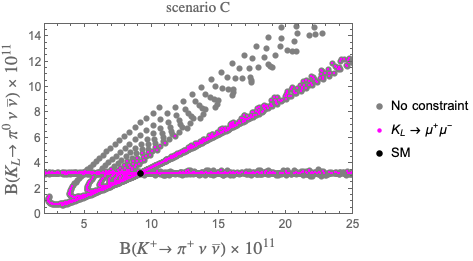}
\caption{\baselineskip 10pt  \small  Scenario C. Correlation between ${\cal B}(K_L \to \pi^0 \nu {\bar \nu})$ and ${\cal B}(K^+ \to \pi^+ \nu \bar \nu)$. The black dot represents the central value of the SM prediction.}\label{fig:KplusvsKLB4}
\end{center}
\end{figure}

Removing the $\Delta F=2$ constraints using the arguments presented in the Appendix~A makes NP effects much larger than in other scenarios.
In this case very large NP effects
in all $\Delta F=1$ observables are possible so that hinted anomalies in
various rare decays of mesons can easily be explained in the absence of
pure leptonic decays. One example is shown in Fig.~\ref{fig:BsvsBdBB4}, where
the branching ratios for $B_{s,d}\to\mu^+\mu^-$ decays are much more modified than in previous scenarios. Another example is shown in Fig.~\ref{fig:KplusvsKLB4},
where NP effects in $\kpn$ and $\klpn$ can be very large. It is also evident
that without the $K_L\to\mu^+\mu^-$ constraint NP effects would be even larger.

\begin{figure}[h]
\begin{center}
\includegraphics[width = 0.47\textwidth]{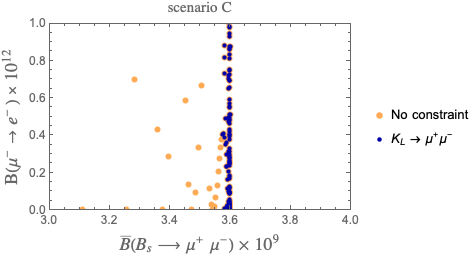} \hskip0.3cm
\includegraphics[width = 0.47\textwidth]{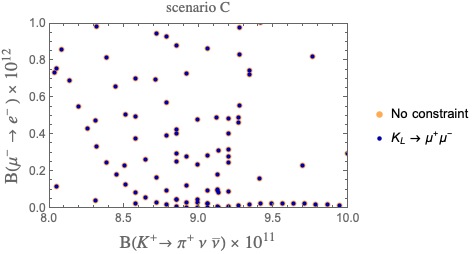}\\
\caption{\baselineskip 10pt  \small
Scenario C. Correlations between ${\cal B}(\mu^-  \to \e^-) $  and  ${\cal \bar B}(B_s \to \mu^+ \mu^-)$ (left panel) and between ${\cal B}(\mu^-  \to \e^-) $  and ${\cal B}(K^+ \to \pi^+ \nu \bar \nu) $ (right panel).}\label{fig:mueconvB4}
\end{center}
\end{figure}

Yet, when the bound from  $\mu-e$ conversion is
taken
into account the imposition of the cancellation of gauge anomalies with the
help of leptons has an important impact on the allowed values of
various observables in the Kaon and $B_{s,d}$ systems.
Indeed the experimental bound on $\mu-e$ conversion eliminates
any significant departures from the SM expectations for $\kpn$ and
$B_{s,d}\to\mu^+\mu^-$. We show this in Fig.~\ref{fig:mueconvB4}.
Similar comments apply to other observables.

\clearpage

\section{Conclusions}\label{sec:conclusions}
We have presented a simple extension of the SM based on the introduction of a new $\text{U(1)}$ gauge group. We have fulfilled the requirement of anomaly cancellation in a rather special way, defining the charges under the new group in terms of  three new parameters $\epsilon_i,\,i=1,2,3$ that are constrained to sum to zero, leaving only two as free parameters. Each of the three parameters governs one of the three fermion generations, so that quark and lepton sectors are  connected to each other as far as their behaviour under the new gauge group is concerned.  This leads to a constrained phenomenological panorama in which quark and lepton observables are correlated with each other. Independently of the structure of the $Z^\prime$ couplings to fermions, that we have varied considering a few scenarios, we have found that  imposing bounds from experimental data relative to $\Delta F=2$  processes to constrain quark favour observables is not enough to prevent large deviations of these from their SM values. However, when constraints on lepton decays are imposed, the room for deviations is much reduced. In practice, quark and lepton sectors act simultaneously to produce a phenomenology similar to the SM. This can be a challenge to experiment but also shows that the pattern of data that we have at our disposal confirms the SM but does not exclude other scenarios like the one we are presenting.

We have also presented a scenario, following \cite{Buras:2014sba}, in which
a suitable hierarchy between LH and RH flavour violating couplings allows to
remove the $\Delta F=2$ constraints. In this case, as illustrated in a number
of plots in Section~\ref{SC}, very large NP effects
in all $\Delta F=1$ observables can be obtained so that possible anomalies in
various rare decays of mesons can easily be explained in the absence of
pure leptonic decays. Yet as discussed on the examples of $\kpn$ and
$B_{s,d}\to\mu^+\mu^-$ even
in this case the imposition of the cancellation of gauge anomalies with the
help of leptons has an important impact on the allowed values of their
branching ratios. This is in particular the case of the experimental bound on $\mu-e$ conversion that eliminates
any significant departures from the SM expectations  of
$B_{s,d}\to\mu^+\mu^-$ and $\kpn$ branching ratios as seen in Fig.~\ref{fig:mueconvB4}.

The main lessons from our paper are the following ones:
\begin{itemize}
\item
  In models in which the gauge anomalies in the quark sector are cancelled
  by the corresponding anomalies in the lepton sector the impact of
  the present experimental bounds on leptonic processes, such as
  $\mu\to e\gamma$, $\mu\to e^-e^+e^-$ and in particular $\mu-e$ conversion
  can be very large unless the model has more free parameters than the
  simple models considered by us.
\item
  It appears then that a strategy for obtaining large NP effects while
  satisfying gauge anomaly cancellations is to construct models
  in which quark and lepton sectors are separately anomaly free.
  \end{itemize}

\section*{Acknowledgements}
We thank Robert Szafron for useful discussions.
J.A. acknowledges financial support from the Swiss National Science Foundation (Project No.  P400P2183838).
The research of A.J.B and M. C-S was supported by the Excellence Cluster ORIGINS,
funded by the Deutsche Forschungsgemeinschaft (DFG, German Research Foundation) under Germany's Excellence Strategy – EXC-2094 – 390783311.
The research of F.D.F. has been carried out within the INFN project (Iniziativa Specifica) QFT-HEP.

\appendix
\boldmath
\section{Removing $\Delta F=2$ Constraints}
\unboldmath
In the presence of both LH and RH
couplings of a $Z^\prime$ gauge boson to SM quarks left-right (LR) $\Delta F=2$ operators are generated whose contributions to the mixing amplitudes $M_{12}^{bq}$ and $M_{12}^{sd}$ in all three mesonic systems are enhanced through renormalisation group effects  relative to left-left (VLL) and right-right (VRR) operators. Moreover in the
case of $M_{12}^{sd}$ additional chiral enhancements of the hadronic matrix elements of LR operators are present. As pointed out in \cite{Buras:2014sba} this fact can be used to suppress
NP contributions to $\varepsilon_K$ and $\Delta M_K$ through some fine-tuning between VLL, VRR and
LR contributions, thereby allowing for larger contributions to $K\to\pi\pi$ amplitudes while satisfying the $\Delta S=2$ constraints. In \cite{Buras:2014zga}
this idea has been generalized to all three meson systems.  While the fine-tuning required in the case of $K\to\pi\pi$ turned out to be rather large, it is  more modest in case of $\Delta B=2$ transitions.

We repeat here briefly the arguments presented in \cite{Buras:2014sba,Buras:2014zga}. Further details can be found in these two papers.

To this end we write the $Z^\prime$ contributions to the mixing amplitudes as follows \cite{Buras:2012jb}:
\be\label{ZpnewK}
(M_{12}^*)_{Z^\prime}^{sd}  =  \frac{(\Delta_L^{sd})^2}{2M_{Z^\prime}^2} \langle \hat Q_1^\text{VLL}(M_{Z^\prime})\rangle^{sd} z_{sd}\,,
\ee
and
\be\label{Zpnewbq}
 (M_{12}^*)_{Z^\prime}^{bq}  =  \frac{(\Delta_L^{bq})^2}{2M_{Z^\prime}^2} \langle \hat Q_1^\text{VLL}(M_{Z^\prime})\rangle^{bq} z_{bq}\,,
\ee
where $z_{sd}$ and $z_{bq}$ are generally complex. We have
\be\label{deltasupp}
 z_{sd}=\left[1+\left(\frac{\Delta_R^{sd}}{\Delta_L^{sd}}\right)^2+2\kappa_{sd}\frac{\Delta_R^{sd}}{\Delta_L^{sd}}\right],
\qquad \kappa_{sd}=\frac{\langle \hat Q_1^\text{LR}(M_{Z^\prime})\rangle^{sd}}{\langle \hat Q_1^\text{VLL}(M_{Z^\prime})\rangle^{sd}}\,,
 \ee
with an analogous expressions for $z_{bq}$. Explicit expressions for the
renormalisation scheme independent hadronic matrix elements and their values  can be found in \cite{Buras:2014zga}.

Now as seen in Table~5 of  \cite{Buras:2014zga}, both $\kappa_{sd}$ and $\kappa_{bq}$ are negative, implying that with the same sign of LH and RH couplings the last term in (\ref{deltasupp}) could suppress the contribution of NP to $\Delta F=2$ processes.
One finds then that for $M_{Z^\prime}\approx 5\tev$ one has
 $\kappa_{sd}\approx -115$ and $\kappa_{bq}\approx - 4.3$ implying that for $z_{sd}$ and $z_{bq}$ to be significantly below unity
the RH couplings must be much smaller than the LH ones.
This in turn implies that the second quadratic term in the expression for $z_{sd}$
in  (\ref{deltasupp}) can be neglected in first approximation, and we obtain the following hierarchy
between LH and RH couplings necessary to suppress NP contributions to $\Delta F=2$ observables:
\be\label{finetuning1}
\frac{\Delta_R^{sd}}{\Delta_L^{sd}}\simeq-\frac{a_{sd}}{2\kappa_{sd}}, \qquad\qquad\qquad \frac{\Delta_R^{bq}}{\Delta_L^{bq}}\simeq-\frac{a_{bq}}{2\kappa_{bq}}\,.
\ee
The parameters $a_{sd}$ and $a_{bq}$ must be close to unity in order to make the suppression effective. How close they should be to unity depends on present and future results for hadronic and CKM parameters in $\Delta F=2$ observables.

Unfortunately the present errors on the hadronic matrix elements are quite large, and do not allow a precise determination of the level of fine-tuning required.
An estimate can be found in Fig.~6 of \cite{Buras:2014zga}.

In any case the fact that  $a_{sd}$ and $a_{bq}$ introduce in each case two
new parameters allows us with some tuning of parameters to weaken the impact
of $\Delta F=2$ constraints on rare decays and even eliminate them which is not
possible in LHS and RHS scenarios.
On the other hand,
due to the hierarchy of couplings and the absence of LR operators in the
rare decays considered by us, rare decays are governed again by LH couplings as in the LHS,
 with the bonus that now the constraint from $\Delta F=2$
 observables can be ignored\footnote{In the case of the correlation of $\klpn$ and $\kpn$ this observation has been made first in  \cite{Blanke:2009pq}, where
   the removal of the correlation of these decays with
   $\varepsilon_K$ in the presence of both LH and RH couplings allowed
   to go beyond the two branch structure as seen by comparing the Figs.~(3) and (7)
   in \cite{Buras:2014zga}.}.  As $\kappa_{sd}\gg\kappa_{bq}$ the
hierarchy of couplings in this scenario
 must be much larger in the $K$ system than in the
 $B_{s,d}$ systems.

 The main message of \cite{Buras:2014sba,Buras:2014zga}
 is the following one: by appropriately
choosing the hierarchy between LH and RH flavour violating $Z^\prime$
couplings to quarks one can eliminate to a large extent the constraints
from $\Delta F=2$ transitions even in the presence of
large CP-violating phases at the price of sizable fine-tuning.
But it should be noted that after the $Z^\prime$ has been integrated out, its LH and RH couplings are scale independent below $M_{Z^\prime}$ and so for a given $M_{Z^\prime}$
this tuning has to be done only once.

The implications of this are rather profound. Even if
in the future the SM would agree perfectly with all $\Delta F=2$ observables,
this would not necessarily
imply that no NP effects can be seen in rare decays. While, in particular in the $K$ system, this requires some severe fine-tuning, we think it is interesting
to consider this possibility.

\bibliographystyle{JHEP}
\bibliography{Bookallrefs}
\end{document}